\documentclass[floatfix]{emulateapj}

\usepackage{natbib}
\usepackage{xspace}
\usepackage[usenames,dvipsnames]{xcolor}
\usepackage[backref,
            breaklinks,
            colorlinks,
            urlcolor=Red,
            linkcolor=Maroon,
            citecolor=Blue]{hyperref}
\usepackage[capitalise,nameinlink]{cleveref}
\usepackage[all]{hypcap}

\newcommand{\keyw}[1]{\textcolor{gray}{#1}}

\begin{document}

\shorttitle{Observations of Solids in Protoplanetary Disks}
\shortauthors{Andrews}

\title{Observations of Solids in Protoplanetary Disks}

\author{Sean M. Andrews}
\affil{Harvard-Smithsonian Center for Astrophysics, 
       60 Garden Street, Cambridge, MA 02138, USA}
\email{sandrews@cfa.harvard.edu}

\begin{abstract}
This review addresses the state of research that employs astronomical (remote 
sensing) observations of solids (``dust") in young circumstellar disks to learn 
about planet formation.  The intention is for it to serve as an accessible, 
introductory, pedagogical resource for junior scientists interested in the 
subject.  After some historical background and a basic observational primer, 
the focus is shifted to the three fundamental topics that broadly define the 
field: (1) {\it demographics} -- the relationships between disk properties and 
the characteristics of their environments and hosts; (2) {\it structure} -- the 
spatial distribution of disk material and its associated physical conditions 
and composition; and (3) {\it evolution} -- the signposts of key changes in 
disk properties, including the growth and migration of solids and the impact of 
dynamical interactions with young planetary systems.  Based on the state of the 
art results in these areas, suggestions are made for potentially fruitful lines 
of work in the near future.
\end{abstract}


\section{Motivation}

The focus on {\it origins}, our cosmic context, is innate to astrophysics.  The 
past two decades have seen remarkable progress and profound, renewed interest 
in the subject, primarily due to advances in the detection and characterization 
of planets around other stars (see \citealt{howard13b} and \citealt{fischer14} 
for recent reviews).  There has been a considerable observational investment in 
studying exoplanet demographics, quantifying their diverse properties 
\citep[masses, orbits; e.g., see the reviews by][]{marcy05,udry07,winn14} and 
identifying some fundamental trends \citep[e.g., with metallicity or host mass; 
see][]{johnson10b}.  The return on that investment is a better understanding of 
the key processes that govern the formation and evolution of planetary 
systems.  The observed properties of the exoplanet population place important 
boundary conditions on theoretical explanations of those processes.  

The analogous study of the protoplanetary disks orbiting young stars offers a 
powerful complement to that work on exoplanets.  Disk observations offer 
insights on the pivotal `initial conditions' available for making planetary 
systems, as well as the early co-evolution and interaction of planets and their 
birth environments.  Ultimately, the observed properties of both disks and 
exoplanets can be employed to inform, test, and refine models of the planet 
formation process (e.g., \citealt{benz14} review a prominent approach to this 
problem).  

The mutual evolution of disks and their planetary systems is terrifically 
complicated.  The best theoretical models include functionally limited physics, 
and do not yet successfully predict the measured properties of exoplanets given 
the inferred parameters of disks.  Much of the lingering uncertainty is due to 
our relative ignorance of disk properties.  Disks are rich in information 
content for planet formation models, but observationally and physically messy.  
Fortunately, recent technical advances are facilitating new, improved disk 
observations that can drive rapid development in those models.

With that bright future in mind, this review is intended as an introductory 
pedagogical resource for the observational side of protoplanetary disk 
research.  The targeted audience is junior members of the community (graduate 
students and postdoctoral scientists).  The content is not comprehensive; the 
focus is primarily on fundamental issues and popular topics related to disk 
{\it solids} (referred to interchangeably as ``dust" or ``particles").  
Although all the key techniques are represented, there is a bias toward 
measurements at radio wavelengths motivated by the recent commissioning of the 
revolutionary Atacama Large Millimeter/submillimeter Array (ALMA).  

The review is organized as follows.  \autoref{sec:history} offers a brief 
historical overview of disk observations, to provide some context for the 
ongoing work in the field.  \autoref{sec:obs} is a primer for interpreting 
observations of disk solids.  
\hyperref[sec:demographics]{Sections~\ref*{sec:demographics}}, 
\ref{sec:structure}, and \ref{sec:evolution} discuss the state of the art 
constraints on the demographics of the disk population, the inferred structures 
and physical conditions in disks, and insights on how those properties change 
with time, respectively.~\autoref{sec:future} summarizes some open questions, 
with an eye toward future observational research.

\section{Historical Background}\label{sec:history}

There is a long history of theoretical interest in protoplanetary disks, 
particularly the precursor to the solar system (the {\it solar nebula}).  
Although there are earlier metaphysical conceptions, the familiar ``nebular 
hypothesis" of a flattened, rotating structure as the origin of the observed 
co-planar planetary orbits was well formulated by \citet{kant1755} and 
\citet{laplace1796} before subsequent elaboration shifted to astronomers (e.g., 
\citealt{moulton1900,moulton1905,chamberlin1900}; see the historical reviews by 
\citealt{brush78a,brush78b,brush81} or \citealt{wood88}).  As more physically 
motivated theories of the star formation process were developed, such a disk 
structure was found to be a natural consequence of gravitational collapse in a 
molecular cloud core endowed with some angular momentum \citep{hoyle60,
cameron62,cassen81,cassen83,terebey84}.  

Meanwhile, taking the existence of such disks as a given, the theoretical study 
of their physical characteristics and evolution was split into two camps; one 
focused on planet formation (see the classic reviews by \citealt{cameron88} and 
\citealt{lissauer93}, or the historical summary of a seminal period by 
\citealt{brush90}), and the other emphasized the processes of accretion and 
momentum transport (see \citealt{pringle81} and \citealt{papaloizou95}).  These 
topics were mature enough to accumulate a vast literature by the 1980s, well 
before many astronomers were completely satisfied with the {\it observational} 
evidence for such disks (let alone extrasolar planetary systems).  

The discovery and early characterization of young stars, the T Tauri (or Orion) 
variables \citep{joy45,joy49,herbig62} and their massive Herbig AeBe 
counterparts \citep{herbig60}, hinted at a close connection with circumstellar 
material.  Influenced by the star formation simulations of \citet{larson69,
larson72}, the blue excesses, line emission \citep[e.g.,][]{strom71,strom72a}, 
and enhanced infrared continua \citep{mendoza66,mendoza68,geisel70,gillett71} 
associated with these young stars were seen as evidence for the circumstellar 
``shells" or envelopes predicted to be remnants of the cloud collapse.  There 
was debate on whether these features were produced solely by hot 
($\sim$10$^4$\,K) gas \citep[e.g.,][]{strom71,strom72a,strom75,strom72b,
rydgren76,warner77}, or if cooler ($\sim$10$^2$\,K) dust was an important 
contributor \citep[e.g.,][]{cohen73,cohen80,cohen79,rydgren81,rydgren83,
rydgren82}.  Ultimately, it was recognized that both components are required; 
an emerging model attributed the hot (blue) excess to accretion shocks at the 
stellar surface \citep{lyndenbell74,uchida84,bertout88} and the cool (red) 
excess to dust on $\sim$AU scales \citep{cohen83,cohen85b,lada84,rydgren84,
rydgren87}.  

That realization was part of a pronounced shift in the mid-1980s away from 
``shells" and toward accretion disks.  Rather than being precipitated by a 
single measurement or study, this conceptual evolution was driven by the 
confluence of several key independent lines of {\it indirect} evidence 
\citep[most of these were clearly summarized in the seminal review 
by][]{shu87}: the observed large (linear) polarizations of T Tauri stars were 
indicative of scattering off of flattened dust structures 
\citep[e.g.,][]{elsasser78,vrba79,bastien82,bastien88}, which were just then 
being resolved in the near-infrared \citep{grasdalen84,beckwith84,strom85}; 
collimated bipolar outflows were being linked to an accretion disk origin 
\citep[e.g., see the review by][]{lada85}; asymmetric (blueshifted) forbidden 
emission line profiles were interpreted as smaller scale winds or outflows that 
are partially obscured by an extended disk \citep[e.g.,][]{edwards87,cabrit90}; 
the outburst behavior in the FU Orionis variables was being associated with 
disk instabilities \citep{hartmann85,lin85b}, forging important links to the 
fields of cataclysmic variables \citep[e.g.,][]{kenyon84,lin85a} and viscous 
accretion disks \citep[e.g.,][]{lyndenbell74}; optically thin mm/radio flux 
measurements confirmed that spherical dust distributions were inconsistent with 
the observed optical extinctions (\citealt{beckwith90}; see also 
\citealt{churchwell87}); and perhaps most influentially, increased access to 
infrared measurements, especially from {\it IRAS} \citep[e.g.,][]{rucinski85,
strom88,harris88,wilking89}, was motivating substantial modeling developments 
to explain the broadband spectral energy distributions (SEDs) in the context of 
irradiated dust disks \citep{adams86,adams87,adams88,kenyon87}.

\begin{figure}[t!]
\epsscale{1.1}
\plotone{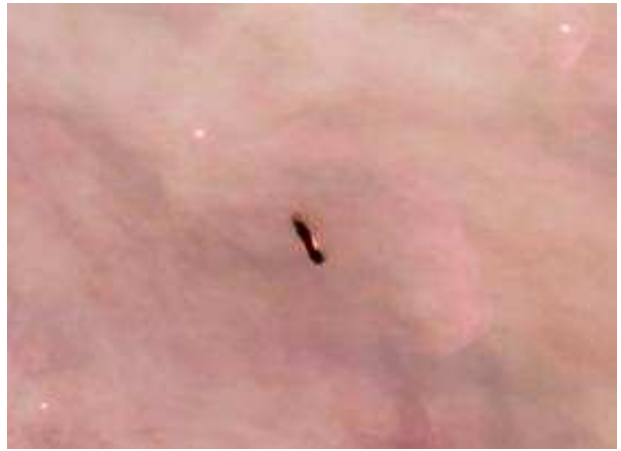}
\figcaption{An optical {\it HST} image of a nearly edge-on disk, seen in 
silhouette against the bright background emission of the Orion Nebula 
\citep{robberto13}.  It was direct images like this that ultimately confirmed 
the theoretical ideas about the ubiquity and basic structures of disks around 
young stars.  \\
\keyw{\it{image credit: NASA, ESA, M.~Robberto (STScI/ESA), and the Hubble 
Space Telescope Orion Treasury Project Team}}  
\label{fig:proplyd}
}
\end{figure}

In the minds of many researchers in the field, this shift in thinking 
culminated in a spectacular ``seeing is believing" confirmation with resolved 
{\it Hubble Space Telescope} ({\it HST}) observations of disks in silhouette 
against the bright background of the Orion Nebula \citep{odell93,odell94,
mccaughrean96}; a favorite example is shown in \Cref{fig:proplyd}.  Around the 
same time, spatially resolved images of scattered light from {\it HST} 
\citep[e.g.,][]{burrows96,stapelfeldt98} and thermal emission from the first 
generation of millimeter-wavelength interferometers \citep[e.g.,][]{sargent87,
sargent91,koerner93,hayashi93,lay94,lay97,koerner95,dutrey94,dutrey96,mundy96,
mannings97} also played key roles in shaping the field.

Since then, the past $\sim$20 years have seen a remarkable proliferation of 
disk measurements that have nurtured the growth of a new, {\it observational} 
branch to the study of planet formation.  It is not unreasonable to say that 
unresolved photometry and spectroscopy from the {\it Spitzer Space Telescope} 
accounts for a substantial share of the maturation of this field, in terms of 
demographics, dust composition and basic structure, rapid development of 
radiative transfer tools, and hints of evolution \citep[e.g.,][]{hillenbrand08,
evans09,furlan09,furlan11}.  But meanwhile, spatially resolved data have been 
playing a steadily increasing role in those advances.  With ambitious new 
technologies being commissioned, spatially resolved observations are poised to 
facilitate transformational progress.

\section{Essentials of Disk Observations} \label{sec:obs}

Before delving into the state of the field and its potential future directions, 
it will be useful to go through a basic primer on disk observations and how 
they are interpreted in the context of disk properties.  

Protoplanetary disks are dense, geometrically flattened circumstellar 
structures composed of a trace population of solids (initially $\sim$1\%\ by 
mass) suspended in a reservoir of (primarily) molecular gas.  Gravity and 
angular momentum conservation ensure that disk densities decrease with radial 
separation from the host star, $r$ \citep[e.g.,][]{terebey84,lyndenbell74}; 
thermal pressure and turbulence determine how the densities fall off with 
height above the midplane, $z$ \citep[e.g.,][]{whipple72,weidenschilling77,
dubrulle95}.  The disk temperatures are almost entirely controlled by the 
passive stellar irradiation of the solids \citep{adams87,kenyon87}, which have 
large broadband opacities that dominate the heating and cooling rates 
\citep{ossenkopf94,pollack94}.  Irradiation heating implies that temperatures 
decrease with $r$.  Starlight intercepted by dust in the disk surface layers 
(atmosphere) is re-radiated out to space and deeper into the disk structure; 
this results in a vertical thermal inversion, where the midplane is cooler than 
the atmosphere \citep{calvet91,chiang97,dalessio98}.  In this basic structural 
framework (see \autoref{sec:structure} for more details), the solids completely 
dominate the opacity budget.  Coupling this with the high sensitivity of 
broadband detectors, it makes sense that observations of the {\it continuum 
emission} generated by solids are the most efficient probes of many disk 
properties. 

That continuum has two primary contributors, from thermal (re-processed stellar 
energy, which dominates) and scattered (reflected starlight) radiation.  The 
specific intensity at frequency $\nu$ along a given line of sight $s$ is 
intimately related to the physical conditions, structural distribution, and 
material properties of the disk solids through the formal radiative transfer 
equation,
\begin{equation}
dI_{\nu} = \rho \,\, \kappa_{\nu} (S_{\nu} - I_{\nu}) \, ds,
\label{eq:RT}
\end{equation}  
where $\rho$ is the density, $\kappa_{\nu}$ is the (absorption+scattering) 
opacity per gram of the solid material, and $S_{\nu}$ is the source function 
(the ratio of emissivity to opacity, with absorption and scattering 
contributions).  The condensed notation of \cref{eq:RT} obscures the tremendous 
complexity of the radiative transfer problem.  A complete solution requires the 
energy input from the host star as well as the three-dimensional distributions 
of $\rho$, $\kappa_{\nu}$ (and the associated {\it directional} scattering 
properties), and $S_{\nu}$.  The major challenge is that $S_{\nu}$ also depends 
on these properties ($\rho$, $\kappa_{\nu}$, etc.).  In practice, this 
complicated feedback is handled numerically, often with Monte Carlo simulations 
\citep{lucy99,bjorkman01} that treat the propagation of photons into the disk, 
the corresponding heating of the solids, and the escape of photons to the 
observer \citep[e.g.,][]{wolf99,whitney03,dullemond04a,robitaille06,pinte06,
min09,robitaille11}.\footnote{There are several Monte Carlo radiative transfer 
codes designed for or amenable to studying disks available as open-source 
software: {\tt \href{http://www.ita.uni-heidelberg.de/~dullemond/software/radmc-3d}{RADMC-3D}}, {\tt \href{http://www.hyperion-rt.org}{HYPERION}}, {\tt \href{http://www.astrophysik.uni-kiel.de/~star}{MC3D}}, and {\tt \href{http://www.michielmin.nl/mcmax/}{MCMAX}} are commonly-used examples (the latter two require 
contact with the developers for download).  Other options (e.g., {\tt MCFOST}) 
are available for collaborative work.}

\subsection{Thermal Emission} \label{subsec:thermal}

Although appreciating the intrinsic complexity of radiative transfer in disks 
is valuable, it offers little intuition for interpreting the data.  But with 
some simplifications, a more pedagogical model that captures the essential 
features of the thermal emission can be constructed.  Consider a disk where 
scattering is negligible, the structure is vertically thin ($S_{\nu} 
\sim$~constant along $s$), and the material is in thermodynamic equilibrium 
($S_{\nu} = B_{\nu}(T)$, the Planck function at the local temperature).  
\cref{eq:RT} can then be recast in terms of the optical depth, defined so that 
$d\tau_{\nu} \equiv \rho \, \kappa_{\nu} ds$, and integrated to give
\begin{equation}
I_{\nu} = B_{\nu}(T) \, (1 - e^{-\tau_{\nu}})
\label{eq:simpleRT}
\end{equation}
(see Ch.~1 of \citealt{rybicki79} for a derivation).  In this toy model, the continuum 
is simply blackbody radiation weighted by the absorbing column.  Optically 
thick emission acts like a thermometer ($I_{\nu} \approx B_{\nu}$ when 
$\tau_{\nu} \gg 1$) for the $\tau$$\sim$1 surface; optically thin emission 
probes the product of temperature, column density ($N$$\equiv$$\int \rho \, 
ds$), and opacity ($I_{\nu} \approx \tau_{\nu} B_{\nu}$ when $\tau_{\nu} \ll 
1$).  

\begin{figure}[t!]
\epsscale{1.20}
\plotone{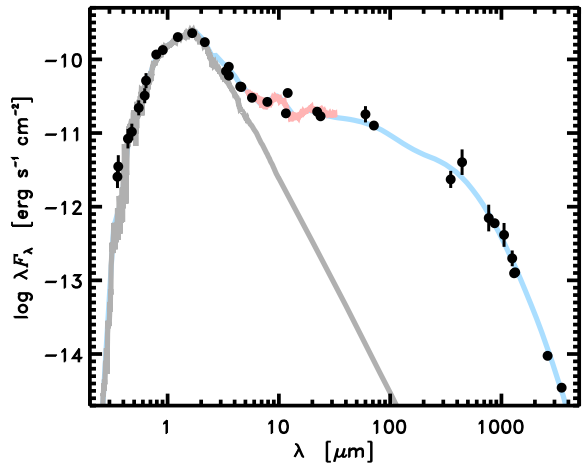}
\figcaption{The SED of the young star GO Tau and its associated disk ({\it 
black} points; data compiled by \citealt{andrews13}).  Models of the stellar 
photosphere ({\it gray}) and an illustrative dust disk ({\it blue}; computed 
for a simple parametric structure using the {\tt RADMC-3D} code) are also shown 
for reference, along with the {\it Spitzer} mid-infrared spectrum \citep[{\it 
red};][]{furlan09}. 
\label{fig:SED_example}}
\end{figure}

The SED is a basic tool for interpreting disk properties\footnote{The SED shows 
how energy, a useful {\it physical} quantity that characterizes the radiative 
transfer of starlight (and intrinsic energy) through the disk material, is 
distributed over frequency or wavelength.  It is defined as $\nu F_{\nu}$ or 
$\lambda F_{\lambda}$, rather than the practical, but physically meaningless, 
{\it measurement} of flux density, $F_{\nu}$ or $F_{\lambda}$ ($=\int I_{\nu} 
\, d\Omega$).  In a begrudging acceptance of standard jargon in the field, the 
latter is referred to as `flux' in this review.}; an example is shown in 
\Cref{fig:SED_example}.  At $\sim$1\,$\mu$m, the SED is dominated by the host 
star.  Thermal emission from the disk starts to outshine the star at 
$\gtrsim$2--5\,$\mu$m, and peaks in the mid- or far-infrared.  This emission is 
optically thick: its luminosity reflects the mean temperature and area of the 
emitting region, and its slope is related to the radial temperature gradient 
(\citealt{adams87,kenyon87}; 
\hyperref[subsec:vertical]{Sect.~\ref*{subsec:vertical}}).  Broad spectral 
features (e.g., \cref{fig:SED_example} near 10\,$\mu$m) provide some insight on 
the dust mineralogy (e.g., \citealt{cohen85a,waelkens96,meeus01,
kessler-silacci05}; \hyperref[subsec:material]{Sect.~\ref*{subsec:material}}).  
The turnover in the 
far-infrared marks the transition to optically thin emission.  The luminosity 
at longer wavelengths scales with the product $\kappa_{\nu} B_{\nu} M_{\rm 
dust}$, where 
$M_{\rm dust}$ is the dust disk mass \citep{weintraub89,beckwith90,adams90}.  The 
mm/radio SED slope probes the {\it shape} of the opacity spectrum 
($\kappa_{\nu}$; \citealt{beckwith91,mannings94}; 
\hyperref[subsec:material]{Sect.~\ref*{subsec:material}}), itself a function of 
the compositions, morphologies, and sizes of the solids 
\citep[e.g.,][]{miyake93,henning96,dalessio01,draine06}.  

\begin{figure}[t!]
\epsscale{1.2}
\plotone{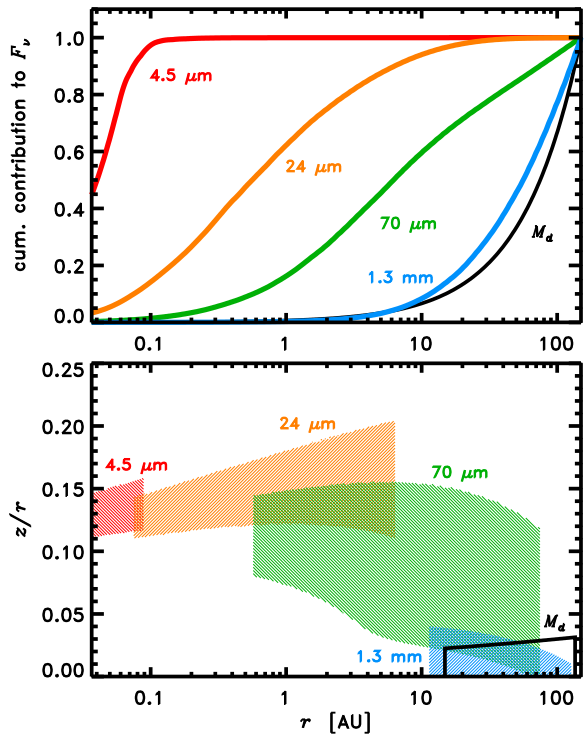}
\figcaption{({\it top}): The cumulative contributions to $F_{\nu}$ at four 
representative wavelengths as a function of $r$ for the SED model in 
\cref{fig:SED_example}.  The cumulative distribution of $M_d$ is also shown (in 
{\it black}).  Note that the 4.5 and 24\,$\mu$m curves are not zero at the 
inner edge, due to the stellar contribution.  ({\it bottom}): A two-dimensional 
map of the regions that emit 80\%\ of the flux (or enclose 0.8\,$M_d$).  
\label{fig:locate}}
\end{figure}

The nature of thermal emission ($\propto$\,$B_{\nu}$) means that shorter 
wavelengths preferentially trace warmer material.  Since higher temperatures 
are usually found closer to the host star, the SED offers a {\it qualitative} 
$\lambda \mapsto r$ mapping (this is slightly complicated when the vertical 
dimension of the disk is considered; see 
\hyperref[subsec:vertical]{Sect.~\ref*{subsec:vertical}}).  \Cref{fig:locate} 
provides a guide to the characteristic emission regions at four representative 
wavelengths, using the SED model in \Cref{fig:SED_example}.  In the 
near-infrared, the emission is produced by both the innermost ($\sim$0.1\,AU) 
part of the disk and the stellar photosphere (in this case with nearly equal 
contributions).  Since the inner disk densities are large and the near-infrared 
opacities are high, this emission is very optically thick; it originates in a 
layer that is a substantial height above the midplane.  Mid-infrared 
emission has similar optical depths, and probes material at large heights 
on $\sim$few AU radial scales.  Far-infrared radiation is emitted over a 
larger range of radii, and can become (partially) optically thin in the outer 
disk ($\sim$tens of AU).  The low optical depths of the mm/radio emission 
offer unique access to cool material in the midplane.

In general, it is not possible to make unique, quantitative inferences about 
disk structures from SEDs alone.  The reason is simple and fundamental.  Disks 
are described by parameters that vary spatially (densities, temperatures) or 
mark specific geometries or locations (e.g., the viewing angle, inner/outer 
boundaries): constraints on such quantities from {\it unresolved} data are 
intrinsically ambiguous (see \citealt{thamm94} or \citealt{chiang99} for 
practical demonstrations).  That is not to claim that SEDs are useless 
diagnostics; indeed, sophisticated SED models have made crucial insights on 
disk properties \citep[e.g.,][]{dalessio98,dalessio99b,dalessio01,dalessio06}.  
But, the ability to alleviate or break some key structural degeneracies {\it 
requires} spatially resolved data.

That said, resolved observations of disks are not trivially obtained.  Most 
nearby disks are located at distances of $\sim$150\,pc: there, the solar system 
dimensions subtend 0\farcs5 on the sky, and AU scales only $\sim$0\farcs01.  A 
quick examination of \Cref{fig:locate} indicates that the thermal emission from 
a typical disk is confined well within the diffraction limits of ground and 
space-based telescopes.  Interferometric measurements are the solution.  
Infrared interferometry offers unique insights on the inner disk structure at 
exquisite angular resolution (for more details, see the comprehensive reviews 
on the subject by \citealt{millan-gabet07} and \citealt{dullemond10}); however, 
the emission at these wavelengths traces a minuscule fraction of the disk mass 
and extent (see \cref{fig:locate}).  Instead, it is (sub-)mm/radio 
interferometry that provides the most robust information on disk structure over 
the widest range of spatial scales.  Contrary to a common misconception, 
observations at these long wavelengths are not solely probing the cool outer 
regions; with sufficient sensitivity and resolution (i.e., from ALMA), they 
will also provide rare access down to $r \sim$few\,AU.  

\begin{figure}[b!]
\epsscale{1.2}
\plotone{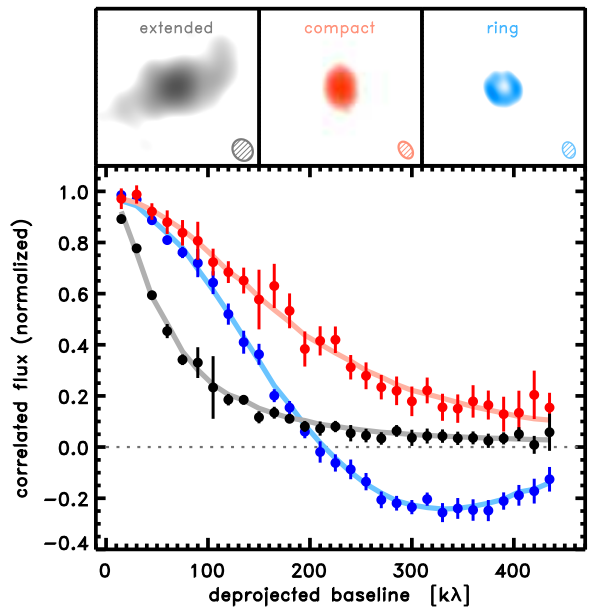}
\figcaption{The 870\,$\mu$m continuum visibility profiles (normalized real 
components, azimuthally averaged accounting for the viewing geometry projected 
onto the sky) for the GSS\,39 ({\it black}), VSSG\,1 ({\it red}) and SR\,21 
({\it blue}) protoplanetary disks \citep{andrews09}.  The corresponding images 
are shown in the top panels, with the synthesized beam dimensions marked in 
their lower right corners.  \label{fig:vis_demo}}
\end{figure}

Interferometers measure the Fourier transform of the brightness distribution at 
a discrete set of spatial scales, determined by the spacings between array 
elements and their projected motion onto the sky plane as the Earth rotates.  A 
complex ``visibility" is measured at each scale; in a coordinate frame with the 
disk origin at the center, the real components are sensitive to the radial 
variation of the brightness and the imaginary components record deviations from 
symmetry.  The data are often displayed in 
condensed format, by azimuthally averaging into one-dimensional profiles that 
represent how the emission depends on spatial scale \citep{lay94,lay97,
hughes07}: \Cref{fig:vis_demo} shows some examples.  These profiles roughly 
correspond to the Fourier transform of the corresponding radial surface 
brightness 
profile.  Recalling some Fourier transform pairs should clarify the behavior: a 
point source in the image domain has a constant amplitude in the visibility 
domain, a compact source has an extended visibility profile (and vice versa), 
and a ring-shaped source (\hyperref[subsec:trans]{Sect.~\ref*{subsec:trans}}) 
makes oscillations like a Bessel function.  The visibilities are usually also 
inverted to construct an image (top panels in \cref{fig:vis_demo}), but any 
analysis should usually be performed in the Fourier domain.

\subsection{Scattered Light} \label{subsec:scatter}

\begin{figure}[t!]
\epsscale{1.1}
\plottwo{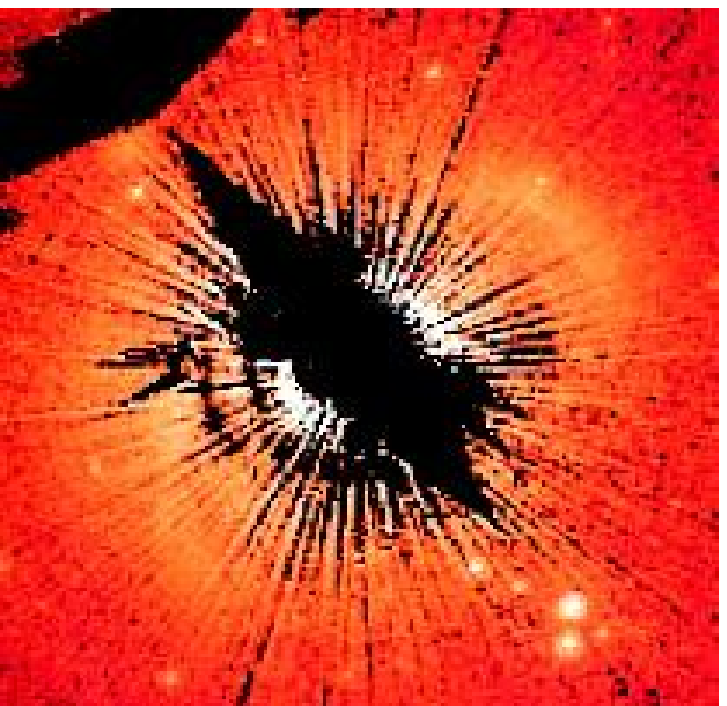}{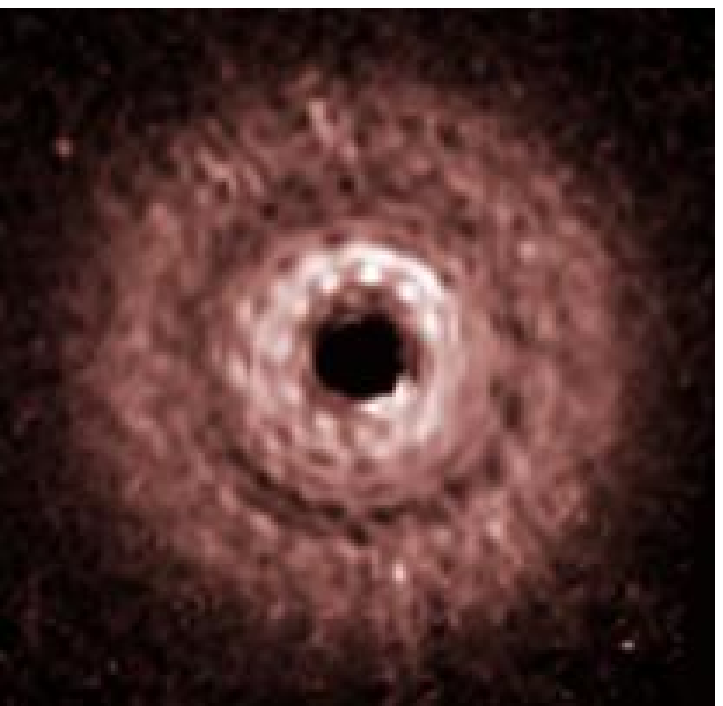}
\figcaption{Coronagraphic images from {\it HST} showing the resolved 
optical and near-infrared starlight scattered off the surfaces of the disks 
around HD\,163296 \citep[{\it left};][]{grady00} and TW\,Hya \citep[{\it 
right};][]{debes13}, respectively.  Some additional examples in different 
contexts can be found in \cref{fig:scat_ims} and 
\hyperref[fig:asym]{\ref*{fig:asym}}.  
\keyw{These images are reproduced with permission from the authors and 
the American Astronomical Society; \copyright~AAS. \\}  
\label{fig:images_example}}
\end{figure}

While the thermal radio emission is extraordinarily informative, it is not the 
only disk probe sensitive to a wide range of spatial scales.  Scattered light, 
primarily in the optical and near-infrared, provides another key resolved 
diagnostic (see \citealt{watson07} for a review); \Cref{fig:images_example} 
shows some examples of scattered light images.  There is no pedagogical 
simplification for the radiative transfer equation when scattering is 
considered, since the source function is coupled to the radiation 
field.\footnote{$S_{\nu}$ includes the integral of $I_{\nu}$ over angle, which 
requires an iterative numerical solution to \cref{eq:RT}.  This is {\it not} 
the case for pure thermal emission, where $S_{\nu}$ depends only on temperature 
(as was assumed in the approximation of 
\hyperref[eq:simpleRT]{Eq.~\ref*{eq:simpleRT}}).}  Starlight is scattered by 
small dust grains with high albedos, having dimensions comparable to the 
spectral peak of the incident radiation field ($\sim$1\,$\mu$m).  The 
corresponding large scattering opacities mean that the observed emission 
originates in the disk surface layers, and can be detected over a large range 
of disk densities (and therefore radial scales).  Such observations are 
fundamental probes of vertical structure 
(\hyperref[subsec:vertical]{Sect.~\ref*{subsec:vertical}}) and can measure 
basic geometric parameters (e.g., inclination).  Incorporating color and 
polarization data constrains the compositions, sizes, and shapes of the 
scatterers (\hyperref[subsec:material]{Sect.~\ref*{subsec:material}}).

The clear advantage of scattered light measurements is that they routinely 
achieve exquisite angular resolution ($\sim$0\farcs1 or less), making them well 
poised to identify and characterize $\sim$10\,AU-scale substructure in disks 
(see \hyperref[subsec:trans]{Sect.~\ref*{subsec:trans}}).  The major challenge 
is contrast, since the host star is typically orders of magnitude brighter at 
the optimal scattering wavelengths.  Coronagraphs or differential imaging 
techniques have been developed to mitigate such issues.  Much of the current 
focus is on polarimetric differential imaging \citep[e.g.,][]{tamura09,
hinkley09,quanz11,perrin14}, which exploits the fact that direct starlight is 
unpolarized, but the scattering by dust grains can induce a significant 
polarization \citep{potter00,kuhn01,perrin04}.  In practice, measurements of 
scattered light near the star are limited by subtraction residuals or 
coronagraph dimensions to $\sim$0.1--0\farcs2 (the ``inner working angle").  At 
larger radii, the data are typically sensitivity-limited because of the diluted 
stellar radiation field (there is a $1/r^2$ depletion of scattering intensity 
on top of any intrinsic fall-off caused by, e.g., a decreasing density of 
scatterers).

\subsection{Synopsis}

\begin{itemize}
\item Solids (dust grains) dominate the disk opacity budget; observations of 
disks are most sensitive to the corresponding bright continuum emission.
\item {\it Thermal} emission is observed over a broad wavelength range; when 
optically thick, it measures the temperature in a surface layer, and when 
optically thin it traces the product of temperature, density, and opacity 
(dust ensemble properties).  
\item The morphology of the infrared SED is roughly related to the radial 
temperature distribution.
\item Spatially resolved data is {\it mandatory} to learn about disk 
structures; mm/radio interferometry measurements uniquely probe material in the 
midplane over a wide range of spatial scales.
\item {\it Scattered} light imaging is a complementary probe of resolved 
structure, especially sensitive to the vertical dimension and sub-structure on 
small scales.
\end{itemize}

\noindent {\it Additional Reading}: the first chapter of \citet{rybicki79}; the 
review of some SED basics by \citet{beckwith99}; the summary reviews on disk 
measurements with mm/radio interferometers by \citet{wilner00} and using 
scattered light images by \citet{mccaughrean00} and \citet{watson07}; the 
overview of disk observations by \citet[][his Ch.~8]{hartmann_book}.

\section{Demographics} \label{sec:demographics}

Planet formation research is still data-limited: the complexity of theoretical 
ideas outpaces observational constraints (although perhaps not for long).  As 
in many similar areas of astrophysics, this means that progress is made in two 
complementary ways: (1) basic demographic studies for large surveys, and (2) 
in-depth analyses of individual (or small groups of) disks.  This section 
focuses on the former.  Case studies and detailed investigations of small 
samples are addressed in 
\hyperref[sec:structure]{Sections~\ref*{sec:structure}} and 
\ref{sec:evolution}.  

Simple and practically accessible observational diagnostics are required to 
accumulate a sample large enough to quantify the statistical distribution of 
some property, or connections between properties, in a disk population.  Key
examples include the use of infrared photometry to describe the basic 
structural evolution of circumstellar material \citep{lada84,adams87,meeus01,
robitaille06,evans09} and the characteristic lifetime of warm dust in disks 
(\citealt{haisch01}; \citealt{hernandez08}; see the summary by 
\citealt{mamajek09}).  Simple accretion diagnostics, like the luminosity of the 
blue excess or the strength of a bright emission line, are also commonly 
employed to probe evolution \citep[e.g.,][]{hartmann98,sicilia-aguilar05} and 
disk-host relationships \citep[e.g.,][]{muzerolle05,natta06}. 

From the perspective of planet formation, the disk {\it mass} ($M_d$) is a 
fundamental demographic parameter: in any theory, the efficiency of planet 
formation depends critically on the amount of raw building material available.  
Moreover, $M_d$ can be estimated from a mm/radio flux measurement ($\propto 
M_{\rm dust}$; \hyperref[subsec:thermal]{Sect.~\ref*{subsec:thermal}}) given 
some assumptions for (or ideally measurements of) the dust-to-gas ratio, dust 
opacity, and temperature.\footnote{For a dust-to-gas mass ratio $\zeta$, the 
total disk mass is $M_d = M_{\rm dust} (1 + \zeta^{-1})$, or $M_d \approx 
M_{\rm dust}/\zeta$ for a typical case where $\zeta \ll 1$.}  This section 
addresses observational insights on the principal issues that affect disk 
masses, specifically the intrinsic relationship with host mass ($M_{\ast}$; 
\hyperref[subsec:mdms]{Sect.~\ref*{subsec:mdms}}) and externally driven 
evolution through interactions with the local 
(\hyperref[subsec:multi]{Sect.~\ref*{subsec:multi}}) and global 
(\hyperref[subsec:enviro]{Sect.~\ref*{subsec:enviro}}) environment.  A brief 
comment on what has been learned about the age-related evolution of the disk 
mass distribution is also made 
(\hyperref[subsec:massevol]{Sect.~\ref*{subsec:massevol}}).

\subsection{Dependence on Host Mass} \label{subsec:mdms}

Most theoretical studies of planet formation assume that $M_d \propto 
M_{\ast}$.  The intuition for this simple scaling relation follows from the 
star formation paradigm \citep[cf.,][]{shu87}, where a protostar accretes mass 
through its disk, which is fed by an envelope reservoir.  The similar shapes of 
the observed core (envelope) and star mass functions affirm this kind of 
``gravitational" scaling \citep[e.g.,][]{motte98,enoch06}, but it need not 
apply to the intermediary disks.  Even if a disk/host mass relationship were 
imprinted at the star formation epoch, it may not persist throughout its 
(potentially $M_{\ast}$-dependent) evolution up to the ages when disks become 
observable and form planets ($\gtrsim 1$\,Myr later).  

And yet, the demographics of the exoplanet population offer compelling evidence 
in favor of such a relationship.  Radial velocity surveys have demonstrated 
that the frequency of giant planets in compact ($\lesssim$~few AU) orbits 
around nearby field stars scales linearly with host mass \citep{johnson07,
johnson10b,bowler10}.  Since the giant planet formation efficiency should 
roughly scale with the disk mass \citep[e.g.,][]{pollack96,hubickyj05}, this 
has been interpreted as strong, indirect evidence for a link between $M_d$ and 
$M_{\ast}$.  

\begin{figure}[t!]
\epsscale{1.15}
\plotone{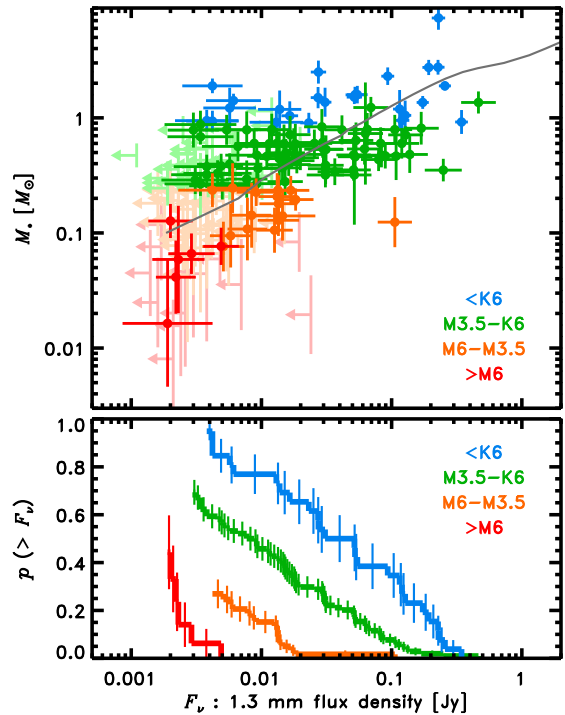}
\figcaption{({\it top}): The 1.3\,mm continuum fluxes ($\propto M_d$) from 
Taurus disks as a function of their host masses \citep[cf.,][]{andrews13}.  
Upper limits on $F_{\nu}$ are shown as faded markings with left-pointing 
arrows.  Despite considerable scatter, there is a clear correlation present 
that is consistent with a linear $M_d \propto M_{\ast}$ scaling; the gray curve 
shows such a relationship for the 2\,Myr \citet{siess00} model isochrone and 
the \citeauthor{andrews13}~assumptions for linking $F_{\nu}$ and $M_d$.  ({\it 
bottom}): The cumulative distributions for $F_{\nu}$ in four different host 
spectral type bins.  These distributions are truncated by the survey 
sensitivity threshold at the low-flux end.
\label{fig:mdms}}
\end{figure}

The signature of that relationship was lacking in early mm/radio continuum 
surveys \citep[e.g.,][]{beckwith90,osterloh95,aw05}, although the sampling in 
$M_{\ast}$ was restricted to around 1\,$M_{\odot}$.  When extended to disks 
orbiting very low-mass hosts ($\sim$0.1\,$M_{\odot}$), significantly fainter 
emission is measured \citep{klein03,scholz06,schaefer09}.  A complete 1.3\,mm 
continuum census across the host mass spectrum in the Taurus region confirms 
that the luminosities increase for hosts with earlier spectral types 
\citep{andrews13}, as demonstrated in \Cref{fig:mdms}.  For a fixed opacity, 
\citeauthor{andrews13}~found reasonable agreement with a linear $M_d \propto 
M_{\ast}$ scaling and a typical disk-to-host mass ratio of 0.2--0.6\%.  That 
said, there is substantial scatter around that relationship, with a 0.7\,dex 
(factor of 5) dispersion in $M_d$ at any given $M_{\ast}$.\footnote{Note that 
this scatter means that some more complex relationships between $M_d$ and 
$M_{\ast}$ (rather than a simple power-law behavior) cannot be easily ruled out 
\citep{andrews13}.}  \citet{mohanty13} independently verify this behavior with 
a different (incomplete, but partially overlapping) sample, and tentatively 
suggested that the relationship flattens (or perhaps even turns over) at high 
$M_{\ast}$.  The same linear scaling is also consistent with the small survey 
of disk masses in the (older) Upper Sco association recently conducted by 
\citet{carpenter14}.

It is tempting to link the observed scaling relations between $M_d$, 
$M_{\ast}$, and giant planet frequency.  Assuming a direct mapping (for solar 
metallicity), and associating the giant planet frequency of 
$\sim$0.07$(M_{\ast}/M_{\odot})$ \citep{johnson10b} with the most massive disk 
progenitors, the Taurus disk mass distribution \citep{andrews13} implies that 
giant planets form when $M_d \gtrsim 0.03$\,$M_{\ast}$.  That mass threshold 
compares well with constraints in the solar system 
(\citealt{weidenschilling77b,hayashi81}; see 
\hyperref[subsec:radial]{Sect.~\ref*{subsec:radial}}), and, coupled with the 
rough linearity of the scalings, is consistent with predictions of the core 
accretion theory for giant planet formation (\citealt{laughlin04}; 
\citealt{ida05}; \citealt{kennedy08}; \citealt{alibert11}; but see 
\citealt{kornet06}).  However, the inferred disk masses are considerably 
uncertain in an {\it absolute} sense (see 
\hyperref[subsec:radial]{Sect.~\ref*{subsec:radial}}--\ref{subsec:material} for 
more details): aside from the notable scatter (\cref{fig:mdms}), the opacities 
and dust-to-gas mass ratio are poorly constrained, and therefore the conversion 
from observed emission ($F_{\nu}$) to $M_d$ may also be biased.  If the 
inferred $M_d$ values are systematically under-estimated (by factors of a few), 
then the alternative disk instability mode of giant planet formation may also 
explain the linear scalings of disk/host/giant exoplanet properties 
\citep{boss11}.

\subsection{Effects of Host Multiplicity} \label{subsec:multi}

Many stars are members of binary or higher-order multiple systems (see the 
recent reviews by \citealt{duchene13} or \citealt{reipurth14}).  The 
multiplicity frequency for Sun-like stars is $\sim$50\%\ in the field 
\citep{abt76,duquennoy91,raghavan10}, and may be even higher in nearby young 
associations \citep{leinert93,ghez93,reipurth93,simon95,kraus11}.  Moreover, 
the distribution of orbital separations in multiples peaks near 100\,AU 
\citep[e.g.,][]{raghavan10,kraus11}, comparable to the typical disk size (see 
\hyperref[subsec:radial]{Sect.~\ref*{subsec:radial}}).  Taken together, these 
properties suggest that the presence of a companion could play a decisive, 
general role in planet formation and the evolution of protoplanetary disks.  

Dynamical interactions in multiple systems alter the structures of the 
associated circumstellar material: theoretical models indicate that tidal 
forces truncate individual disks at $r \gtrsim0.2$--0.5$a$, where $a$ is the 
orbital separation, and clear the inner regions of circum{\it binary} disks at 
$r \lesssim2$--5$a$ \citep[e.g.,][]{lin93,artymowicz94,artymowicz96}.  Despite 
such a disruptive environment, the mature counterparts of these young multiples are known to host planets \citep[e.g.,][]{patience02,raghavan06,desidera07,
bonavita07}.  Observations of the disks in young stellar pairs offer key 
insights on how planet formation is affected in multiple host environments.

\begin{figure}[t!]
\epsscale{1.15}
\plotone{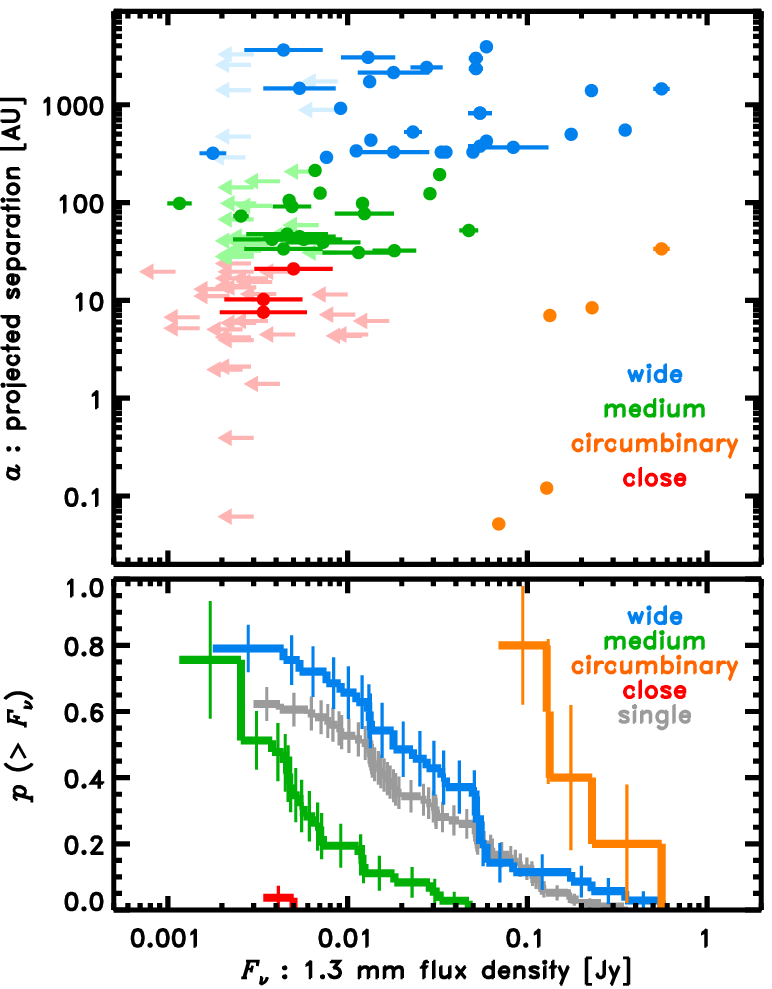}
\figcaption{({\it top}): The 1.3\,mm flux ($\propto M_d$) from disks in stellar 
(spectral type $\le$M5) multiple systems as a function of the separation 
between host pairs (cf., \citealt{harris12}; includes data from 
\citealt{akeson14}).  Symbols are as in \cref{fig:mdms}.  The flux distribution 
shifts to fainter values when the disks are hosted by closer pairs, although 
there is a bimodal distribution at the closest separations where circumbinary 
disks ({\it orange}) can be especially bright (massive).  ({\it bottom}): The 
cumulative distributions for $F_{\nu}$ in different pair separation bins.  The 
gray distribution is for singles.
\label{fig:mdsep}}
\end{figure}

Given the predicted outcomes of these star--disk interactions, the 
observational focus has been on characterizing how disk properties vary with 
$a$ (or its sky-projected equivalent).  Seminal early work by \citet{jensen94,
jensen96} indicated that $M_d$ is diminished for stellar pairs with $a \lesssim 
50$--100\,AU, in quantitative agreement with the expectations of tidal 
truncation models \citep[see also][]{osterloh95,aw05,cieza09}.  Recent 
extensions of those studies with improved sensitivity confirm these results 
\citep{harris12,akeson14}: \Cref{fig:mdsep} illustrates this behavior in the 
Taurus region.  The closest pairs have lower disk masses (but see below) than 
those with wide separations or their singleton counterparts; the median $M_d$ 
scales up by a factor of $\sim$three per decade in $a$.  Similar conclusions 
are drawn from infrared excesses \citep[e.g.,][]{cieza09,
kraus12}.\footnote{However, as was highlighted in 
\hyperref[subsec:thermal]{Sect.~\ref*{subsec:thermal}}, the high optical depths 
at such wavelengths makes it difficult to link the results to mass depletion.  
Indeed, some studies find examples of infrared excess emission regardless of 
the projected host separation \citep[e.g.,][]{white01,mccabe06,pascucci08}.}  
The frequency of short-period giant planets with binary hosts is also 
suppressed \citep{wang14}, consistent with this inferred disk mass depletion 
present at the planet formation epoch.  

However, the separation-dependent pair demographics tell only part of the 
story.  Component-resolved mm observations indicate that the primary disk 
almost always dominates the $M_d$ budget, and that often the disk mass ratios 
are much different than would be expected from the standard theory given the 
stellar masses and separations \citep{jensen03,patience08,harris12,akeson14}.  
Moreover, the resolved sizes of {\it individual} disks in such systems show at 
best marginal consistency with theoretical predictions for their tidal 
truncation radii \citep{harris12}.  Aside from these apparent discrepancies, 
the pair demographics themselves highlight an additional (often unappreciated) 
mystery: unlike any other separation scale, there is a clear bimodal $M_d$ 
distribution for the closest ($< 10$\,AU) pairs (see \cref{fig:mdsep}), with 
most remaining undetected but a few having notably massive circumbinary disks.  

Some of these observations might best be explained in terms of how the process 
of multiple star formation directs mass from the natal envelope to specific 
components in the system \citep[i.e., initial conditions;][]{bate97,bate00,
ochi05}.  But there are also indications that the standard tidal interaction 
models are insufficient; modern hydrodynamic simulations suggest that 
asymmetries (like eccentricities or warps) are fundamental 
\citep[e.g.,][]{kley08,paardekooper08,marzari09}.  Particularly striking are 
the systems with disks that are misaligned with respect to each other 
\citep{jensen04,mccabe11,jensen14,williams14} and/or the orbital planes of 
their respective hosts \citep[e.g.,][]{akeson07,verrier08,andrews10}.

\subsection{Environmental Impact} \label{subsec:enviro}

Most disk observations are necessarily focused on the {\it nearest} 
star-forming regions, which have low stellar densities and few (if any) massive 
stars.  However, most stars form in the significantly different environments of 
dense clusters \citep{lada03,porras03}, where much larger stellar populations 
create opportunities for disks to be affected by two key {\it external} 
environmental factors.  First, high stellar densities increase the probability 
for ``fly-by" interactions, and thereby the tidal disruption of disks 
\citep[e.g.,][]{clarke93,korycansky95,larwood97,boffin98,kobayashi01}.  Second, 
the high-mass tail of the $M_{\ast}$ distribution is populated; disks in the 
immediate vicinity of massive stars will be depleted by photoevaporation, due 
to the locally intense radiation field \citep[e.g.,][]{hollenbach94,
johnstone98}.

There is not yet any clear evidence indicating that tidal stripping from close 
stellar encounters plays a significant role in setting disk properties in rich 
clusters.  Indeed, the expected interaction rate in the nearest massive cluster 
(the Orion Nebula cluster, or ONC) is quite small \citep[e.g.,][]{adams06,
proszkow09}.  However, the relevant observational signature -- a substantial 
radial truncation, and thereby mass depletion \citep[e.g.,][]{breslau14} -- is 
not yet readily available in the ONC or other, more distant, 
regions.

\begin{figure}[t!]
\epsscale{1.15}
\plotone{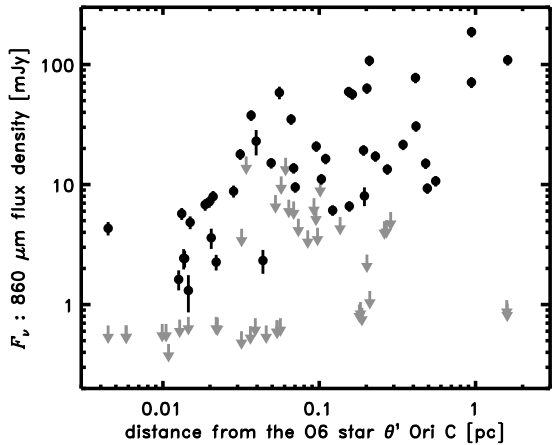}
\figcaption{The 860\,$\mu$m fluxes ($\propto$$M_d$) from disks in the ONC as a 
function of their projected separation from the massive star $\theta^1$\,Ori\,C 
\citep[cf.,][]{mann10,mann14}.  Upper limits (3\,$\sigma$) are marked with gray 
arrows.  Disk masses drop precipitously when their hosts are located in the 
small region ($\lesssim$0.03\,pc) where extreme UV emission from the Trapezium 
stars dominates the radiation field, as expected from photoevaporation models.  
\label{fig:mddob}}
\end{figure}

There is a wealth of data that probes the photoevaporation of disk material 
(around Sun-like hosts) in the proximity of massive stars, typically in the 
form of ionized (shocked) shell structures associated with silhouette disks 
\citep[e.g.,][]{churchwell87,bally98,bally00,henney99,smith03}.  Early attempts 
to quantify the mass depletion in these disks with mm interferometry were 
difficult \citep[e.g.,][]{mundy95,bally98b,eisner06}, but eventually bore fruit 
at higher frequencies \citep{williams05,eisner08}.  \citet{mann09,mann10} 
identified a trend toward smaller disk masses for hosts located closer to the 
OB stars in the center of the ONC.  \Cref{fig:mddob} shows a recent 
confirmation of this trend, which provides crucial quantitative evidence for 
photoevaporation-driven mass loss \citep{mann14}: stars located inside the 
region where energetic ultraviolet photons from the O8 star $\theta^1$\,Ori\,C 
dominate the radiation field ($\lesssim$~0.03\,pc) host only faint (low-mass) 
disks.

\subsection{Mass Evolution} \label{subsec:massevol}

By the time they are observable, disks have already undergone some (perhaps 
substantial) evolution.  The key signatures of that evolution will be covered 
in \autoref{sec:evolution}, but the relevant {\it timescales} are best 
considered from a demographic perspective.  The standard astronomical approach 
tracks how a given property changes between large samples that have different 
mean ages.\footnote{The ages of young stars are not well determined 
\citep[see][]{soderblom14}.  The benchmarks common in the literature and 
propagated here can be interpreted with considerable liberty.}  Application of 
this technique to simple disk tracers, like the infrared excess 
\citep[e.g.,][]{haisch01,hernandez07} or accretion rate indicators 
\citep[e.g.,][]{sicilia-aguilar10}, suggest that the typical inner disk 
($<1$\,AU) survives over a $\sim$5--10\,Myr timescale \citep[but 
see][]{pfalzner14}.  But such constraints offer little in terms of assessing 
the available timeframe for planet formation.  For that, it is critical to know 
the mass depletion rate, or how the $M_d$ distribution changes with time.

The current observational information on $M_d$ evolution is very limited.  Only 
the Taurus region ($\sim$2--3\,Myr) has a mm continuum census complete to a 
sufficiently deep limit \citep{beckwith90,osterloh95,aw05,andrews13}.  
Continuum surveys are also available for the NGC 2024 
\citep[1\,Myr;][]{eisner03,mann15}, Ophiuchus \citep[1--2\,Myr;][]{andre94,
nurnberger98,aw07b}, ONC \citep[1--3\,Myr; e.g.,][]{mann14}, Lupus 
\citep[1--3\,Myr;][]{nurnberger97}, Chamaeleon \citep[1--3\,Myr;][]{henning94}, 
MBM 12 \citep[1--3\,Myr;][]{hogerheijde02}, IC 348 
\citep[3\,Myr;][]{carpenter02,lee11}, $\sigma$ Ori 
\citep[3--5\,Myr;][]{williams13}, $\lambda$ Ori \citep[5\,Myr;][]{ansdell15}, 
and Upper Sco \citep[5--10\,Myr;][]{mathews12,carpenter14} clusters, although 
with inhomogeneous sizes, completeness levels, and sensitivities.  Comparisons 
between these samples are difficult, due to the strong selection effects 
related to the demographic trends discussed above; biased, small samples can 
mimic or obscure real evolutionary changes in the $M_d$ distribution 
\citep{andrews13}.  That said, the low detection rates in the oldest samples 
that have been probed ($\sigma$ Ori, $\lambda$ Ori, and Upper Sco, at 
$\sim$5\,Myr) do indicate an overall decrease in the mean disk mass.

\subsection{Synopsis}

\begin{itemize}
\item Disk mass is the fundamental aggregate property in planet formation 
models.  The factors that influence $M_d$ can be studied demographically with 
large mm/radio continuum photometry surveys, although are subject to uncertainties in the assumed dust-to-gas ratio, opacities, and temperatures.
\item Disk masses are related to their stellar host masses; a roughly linear 
$M_d \propto M_{\ast}$ scaling seems appropriate, although there is a large 
dispersion (0.7\,dex) in $M_d$ for any given $M_{\ast}$.  For standard 
assumptions, the median disk-to-star mass ratio is $\sim$0.2--0.6\%.
\item In multiple star systems, close pairs tend to host low-mass disks.  Most 
of the circumstellar material is usually concentrated around the primary star.  
\item Disk masses are depleted (by photoevaporation) in the immediate vicinity 
of massive (OB) stars.
\item The timescales over which the $M_d$ distribution changes are not yet 
clear, due to incomplete samples and selection effects.  Preliminary results 
suggest that there is significant depletion within 5\,Myr.  
\end{itemize}

\noindent {\it Additional Reading}: the general review by \citet{williams11}; 
the topical reviews of disk dissipation timescales by \citet{hillenbrand08b} 
and \citet{mamajek09}.

\section{Structure} \label{sec:structure}

Although demographic studies reveal some fundamental properties of the disk 
population, their reliance on easy-to-observe ``compound" diagnostics means 
that they cannot tell the whole story.  An important complement is found in 
detailed probes of individual disk structures, which comprise a set of more 
``elemental" measurements.  The single most valuable of these is the {\it 
spatial distribution of mass}.  The efficiency of planet formation is directly 
linked to local disk densities: there must be enough stuff in the right places 
(for enough time) to assemble a planetary system from its progenitor (disk) 
material.  Such a density threshold {\it in solids} is especially important in 
the formation models of terrestrial planets \citep[e.g.,][]{raymond04,kenyon06,
kokubo06} and giant planet cores \citep[e.g.,][]{mizuno80,pollack96,
hubickyj05}.

Inferring the disk structure from its associated observational tracers is 
complicated: there are elaborate dependences between the densities, 
temperatures, and material properties.  Fortunately, different kinds of data 
are especially sensitive to specific aspects of disk structure 
(\hyperref[sec:obs]{Sect.~\ref*{sec:obs}}).  When complementary tracers are 
considered together, they can be used to forge crucial benchmarks for models of 
planet formation and disk evolution.  This section covers the key aspects of 
disk structures, with emphasis on the practical connections between physical 
parameters and data.  It highlights the role of vertical structure in 
regulating the thermal budget 
(\hyperref[subsec:vertical]{Sect.~\ref*{subsec:vertical}}), the radial 
distribution of mass (\hyperref[subsec:radial]{Sect.~\ref*{subsec:radial}}), 
and the characteristic properties of the disk solids 
(\hyperref[subsec:material]{Sect.~\ref*{subsec:material}}).

\subsection{Vertical Structure} \label{subsec:vertical}

\subsubsection{Physical Overview}

The aspect ratio of a rotationally flattened disk is small, $z/r \sim 
\mathcal{O}(0.1)$, since material is gravitationally concentrated at the 
midplane.  But even this small vertical extent has important consequences.  The 
most common model for a disk structure is based on some simple physical 
arguments for the {\it gas}, implicitly assuming that the solids follow the 
same distribution (but see 
\hyperref[subsec:solidevol]{Sect.~\ref*{subsec:solidevol}} for some important 
caveats).  The underlying principle is that the disk is in {\it vertical} 
hydrostatic equilibrium,
\begin{equation}
\frac{\partial P}{\partial z} = -\rho \, g_z,
\label{eq:hse}
\end{equation}
where $P$ is the gas pressure and $g_z$ is the $z$-component of the 
gravitational acceleration.  In most cases, the ideal gas equation of state is 
appropriate and the stellar host dominates the potential.  \cref{eq:hse} is 
then equivalent to
\begin{equation}
\frac{\partial \ln \rho}{\partial z} = -\left[\frac{\mu m_{\rm H}}{k T} 
\frac{GM_{\ast} \, z}{(r^2+z^2)^{3/2}} + \frac{\partial \ln T}{\partial z} 
\right],
\label{eq:hse2}
\end{equation}
with $G$ the gravitational constant, $\mu$ the mean molecular weight, 
$m_{\rm H}$ the mass of a hydrogen atom, and $k$ the Boltzmann constant.  
\cref{eq:hse2} is generally solved numerically, although some common 
simplifications enable an illustrative analytic solution.  For a geometrically 
thin disk ($z \ll r$) with a small temperature gradient ($\partial T/\partial z 
\approx 0$), the solution to \cref{eq:hse2} is a simple Gaussian distribution,
\begin{equation} 
\rho = \frac{\Sigma}{\sqrt{2\pi} H} \exp{\left[-\frac{1}{2} 
\left(\frac{z}{H}\right)^2\right]},
\label{eq:density}
\end{equation}
with
\begin{equation}
H = \frac{c_s}{\Omega} = \left(\frac{k T}{\mu m_{\rm H}} 
\frac{r^3}{G M_{\ast}}\right)^{1/2}
\label{eq:scaleheight}
\end{equation}
denoting a characteristic scale height \citep[the ratio of the sound speed, 
$c_s$, to the Keplerian angular speed, $\Omega$; cf.,][]{shakura73} and 
$\Sigma$ representing a surface density (a boundary condition of the 
integration).

\begin{figure}[t!]
\epsscale{1.2}
\plotone{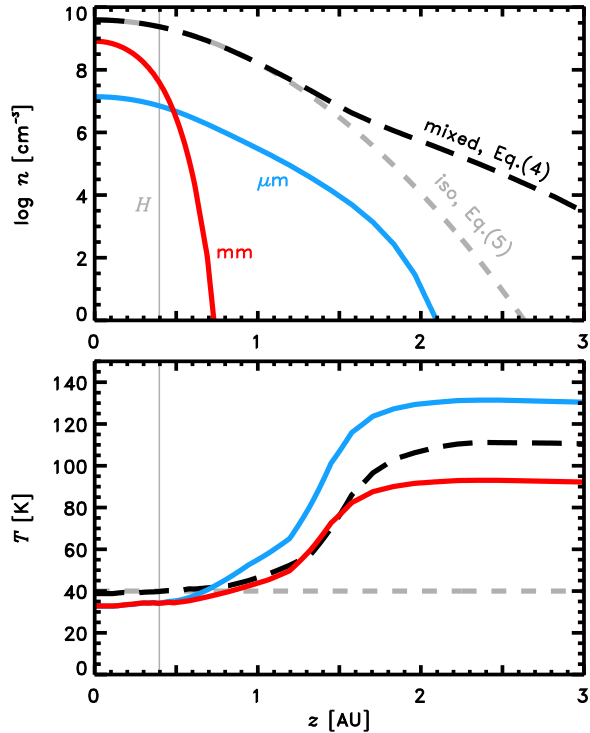}
\figcaption{Vertical cuts at a fixed radius ($r = 10$\,AU) of the density ({\it 
top}) and temperature ({\it bottom}) profiles for three illustrative disk 
structures.  The simple vertically isothermal behavior described in 
\cref{eq:density} is shown as gray dash-dot curves.  The nominal scale height 
(corresponding to the midplane temperature), $H \approx 0.4$\,AU, is marked 
with a vertical gray line.  A more realistic variant, where particles with a 
power-law size distribution are well-mixed with the gas and the vertical 
structure has been iterated using \cref{eq:hse2} and a continuum radiative 
transfer code is shown as dashed black curves.  Finally, the blue and red 
curves represent particles with 1\,$\mu$m and 1\,mm sizes (respectively) from a 
similar model where the well-mixed assumption is relaxed: the solids are 
distributed according to a size-dependent balance between turbulent mixing and 
hydrostatic support, following the prescription of \citet{dubrulle95} (see 
\hyperref[subsec:solidevol]{Sect.~\ref*{subsec:solidevol}}).  All models use 
the same $\Sigma$ and material properties, and employ the {\tt RADMC-3D} code. 
\label{fig:atm_structures}}
\end{figure}

Simple energy arguments based on their observed bolometric luminosities 
indicate that disks are primarily heated by stellar irradiation \citep{adams86,
adams87}.  With their large, broadband opacities maximized near the peak of the 
stellar spectrum, small ($\sim$$\mu$m-sized) grains are the most efficient 
conduit for that irradiation energy.  Starlight absorbed in the disk surface 
layers heats those grains, which then re-emit some of that energy deeper into 
the disk interior to warm the midplane \citep{calvet91,calvet92,malbet91}.  
This {\it external} deposition of energy produces a thermal inversion in the 
disk atmosphere ($T$ increases with $z$) and modifies the vertical density 
distribution (cf., \hyperref[eq:hse2]{Eq.~\ref*{eq:hse2}}; \citealt{chiang97,
chiang99,dalessio98,dalessio99b,bell97,bell99,dullemond01,dullemond02,
malbet01}).  All else being equal, the heating depends on the irradiated area 
of the surface layer.  If $H(r)/r$ is increasing, the tilt of this layer toward 
the star increases that area: such {\it flared} disks intercept more starlight 
than their flat counterparts, and therefore have warmer temperatures \citep[at 
a given $r$;][]{kenyon87,calvet91,chiang97}.  \Cref{fig:atm_structures} 
illustrates some examples of the relationship between the vertical 
distributions of $\rho$ and $T$.

Although it is a simplification, \cref{eq:density} faithfully highlights this
fundamental point about disk structures: the {\it densities and temperatures 
are physically coupled}.  The intrinsic complexity of this coupling is that 
these physical conditions are linked to the radiative transfer of energy 
through the disk, which itself both depends on the densities and sets the 
temperatures.  This feedback between structure and radiative transfer is a 
{\it generic} feature.  

\subsubsection{Observational Constraints}

\begin{figure}[t!]
\epsscale{1.15}
\plotone{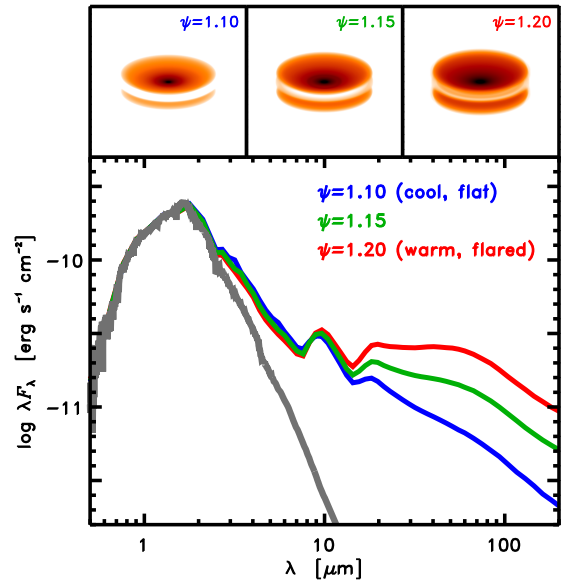}
\figcaption{A demonstration of the impact that the characteristic height has on 
key observables, including scattered light ({\it top}) and the SED ({\it 
bottom}), using a simple structure model and the {\tt RADMC-3D} code.  The 
models have structures described by \cref{eq:density}, with identical 
parameters aside from the shape of the radial $H$ profile.  Each $H(r)$ is 
normalized at 1\,AU (to $\sim$0.04\,AU), and has a power-law scaling $H(r) 
\propto r^{\psi}$ with $\psi = 1.10$, 1.15, and 1.20.  The most flared model 
($\psi = 1.20$) exhibits brighter mid-infrared emission and has a more 
prominent scattered light disk.  
\label{fig:atm_obs}}
\end{figure}

The most common approach used to measure the vertical structure of a disk 
employs the infrared SED as a diagnostic (cf., \cref{fig:locate}), and relies 
on modeling the connection between the height and temperature of the effective 
surface layer \citep[the infrared ``photosphere"; 
e.g.,][]{kenyon87,chiang97,dalessio98}.  Since more energy is absorbed for a 
disk with a larger vertical extent, the temperature of the surface layer -- and 
therefore the luminosity -- scales with its characteristic height.  The rough 
$\lambda \mapsto r$ mapping of the infrared SED (see 
\hyperref[subsec:thermal]{Sec.~\ref*{subsec:thermal}}) means that the spectral 
slope reflects the radial variation of the surface layer height: more flaring 
makes a redder SED.  Some examples of this effect are shown in 
\Cref{fig:atm_obs}.  However, although the SED does depend on the vertical 
structure, it alone cannot be used to unambiguously quantify any associated
physical parameters (e.g., $H$): again, unresolved tracers do not robustly 
measure {\it spatial} properties.  

\begin{figure}[t!]
\begin{center}$
\begin{array}{cc}
\includegraphics[width=1.55in]{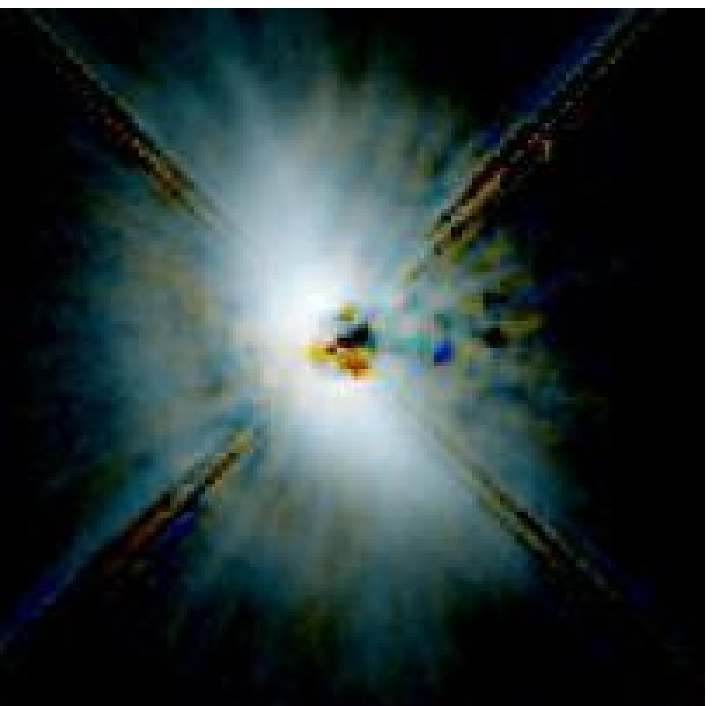} &
\includegraphics[width=1.55in]{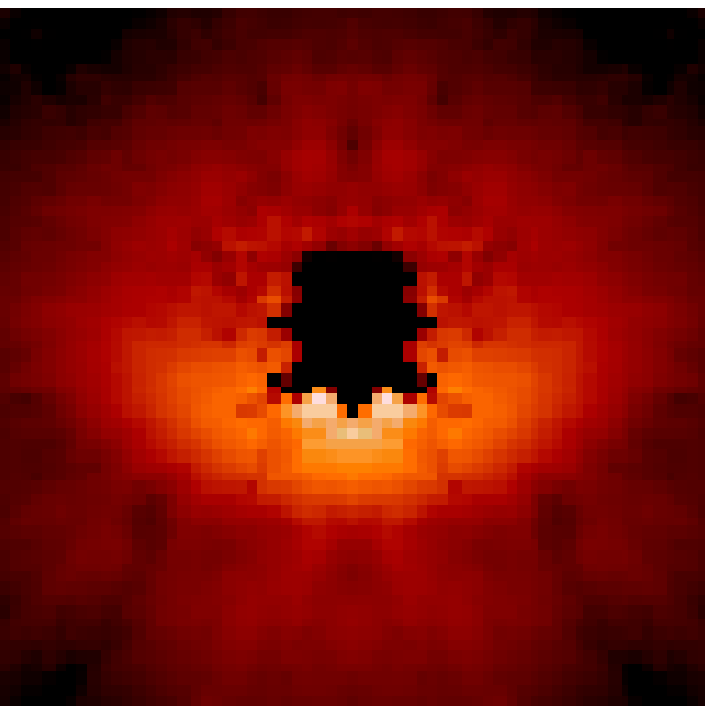} \\
\includegraphics[width=1.55in]{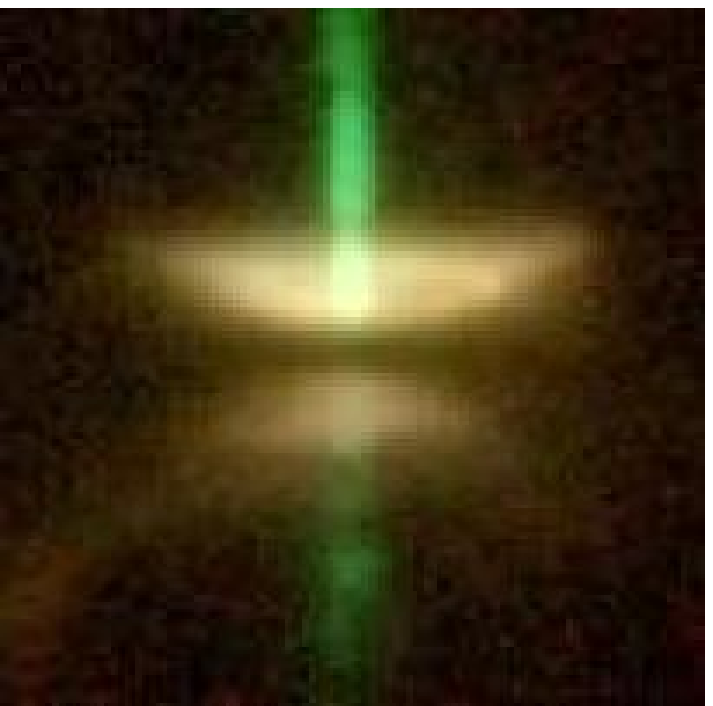} &
\includegraphics[width=1.55in]{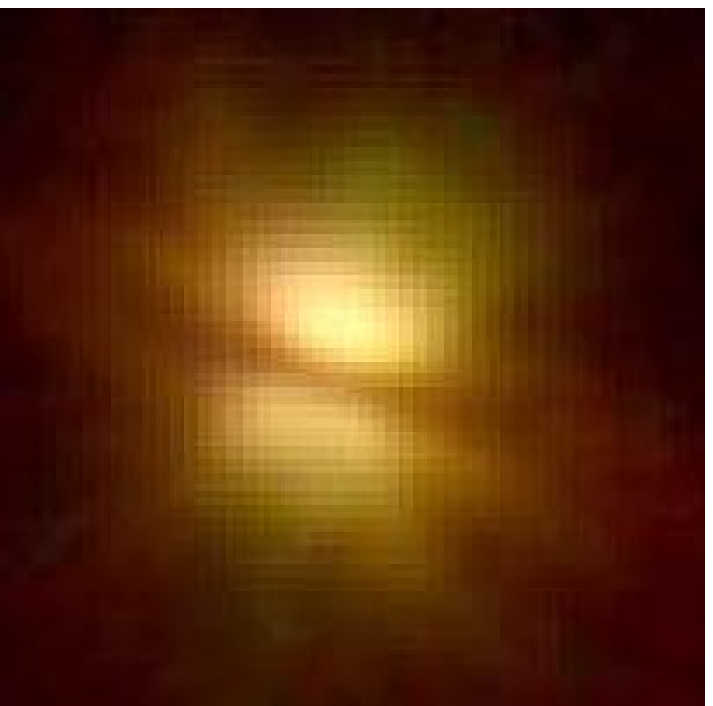} 
\end{array}$
\end{center}
\figcaption{Examples of scattered light morphologies: 
({\it clockwise, from top left}) optical/near-infrared {\it HST} images of GM 
Aur \citep{schneider03}, IM Lup \citep{pinte08}, HV Tau C 
\citep{stapelfeldt03}, and HH 30 IRS (\citealt{burrows96,watson04,
watson07a}; image courtesy of K.~R.~Stapelfeldt). 
\keyw{These images are reproduced with permission from the authors and the AAS 
($\copyright$ AAS) and {\it Astronomy \& Astrophysics} ($\copyright$ ESO).}
\label{fig:scat_ims}}
\end{figure}

The most valuable complement is a resolved map of scattered light emission 
(\hyperref[subsec:scatter]{Sect.~\ref*{subsec:scatter}}).  This emission 
generally appears as a bipolar pair of conical nebulae (representing the 
scattering layers on each side of the disk) with a dark waist (the disk 
interior); some representative examples are shown in \Cref{fig:scat_ims}.  The 
curvature, or opening angle, of these nebulae is set by the shape of the 
scattering surface and thereby the flaring geometry of the disk structure 
\citep[e.g.,][]{bastien90,lazareff90,whitney92,wood98,dalessio99b,takami14}.  
All else being equal, more flaring produces more curvature, larger nebulae, and 
a higher overall luminosity and polarization fraction for the scattered 
light.  The images in \Cref{fig:scat_ims} illustrate some of these effects.  

In practice, the inference of vertical structure parameters from such data 
relies on some assumptions (or external constraints) about the viewing geometry 
and optical properties of the scatterers.  For intermediate inclination angles, 
these degeneracies are usually explored in a restricted forward-modeling 
process \citep{roddier96,close98,grady99,grady00,grady07,potter00,augereau01,
krist02,krist05,schneider03,pinte08,kusakabe12}.  Such ambiguities are 
diminished for extreme inclinations: the variation of the surface layer height 
with radius can be determined most directly from the nebular curvature in the 
edge-on case \citep[e.g.,][]{burrows96,lucas97,krist98,padgett99,stapelfeldt98,
stapelfeldt03,cotera01,wolf03,perrin06,glauser08}, or more indirectly from the 
morphology of the radial brightness profile for face-on disks 
\citep[e.g.,][]{krist00,trilling01,debes13}.  

Using \cref{eq:density} as a parametric structure, models of resolved scattered 
light data indicate $H/r \approx 0.05$--0.25, with only a modest radial 
variation ($\propto r^{0.1-0.3}$).  Similar estimates are made from the 
infrared SED.  As a point of reference, the hydrostatic solution for a 
vertically isothermal disk predicts that $H/r \propto r^{1/4}$.  It is 
worth a reminder that these are not actually {\it measurements} of the pressure scale height $H$, but rather inferences based on 
constraints of the characteristic height of the surface layer where dust 
scatters or absorbs starlight (which are not necessarily co-located).  Although 
these are expected to roughly track one another, there are some additional 
effects that impact their relationship 
(\hyperref[subsec:solidevol]{Sect.~\ref*{subsec:solidevol}}).

One such complication occurs when the characteristic height does not rise 
monotonically with $r$.  If material closer to the star is distributed to 
sufficiently large heights, it can attenuate the radiation field at larger 
radii.  This {\it shadowing} is usually attributed to a ``puffed-up" inner disk 
edge (or {\it wall}), where internal energy generated by viscous dissipation 
(for disks with high accretion rates; e.g., \citealt{dalessio98}) or direct 
irradiation of the inner disk edge \citep{natta01,dullemond01,dalessio05,
calvet05} heats the material on sub-AU scales and inflates the local scale 
height.  However, it may also be produced if vertical mixing is more vigorous 
in the inner disk \citep[e.g.,][]{dullemond04a} or in the presence of 
small-scale substructures.  Regardless of its origin, such obscuration makes 
the outer disk cooler, lowers the effective surface layer, and therefore 
reduces the infrared SED (particularly at longer $\lambda$; a vertically 
enhanced inner edge actually produces {\it more} near-infrared emission) and 
scattered light luminosity \citep{dullemond01,dullemond04a}.  These effects 
have been inferred from SED morphologies \citep[e.g.,][]{meeus01,natta01,
dominik03,acke04}.  A more striking, direct confirmation is found in scattered 
light images that show the flared outer disk emerging out of a shadow at large 
$r$ \citep[e.g.,][]{garufi14b}.

Synoptic infrared photometry surveys are suggesting that small-scale vertical 
substructures may be common.  Variability at some level is ubiquitous 
\citep[see][]{cody14}, but the timescales observed in the mid-infrared are 
surprising: they are much shorter than expected for the disk regions being 
traced ($\sim$few AU; see \cref{fig:locate}; \citealt{sitko08,
morales-calderon09,flaherty12,kospal12,rebull14}).  When near-infrared 
monitoring is also available, a ``see-sawing" wavelength dependence has been 
observed, where the blue and red ends of the spectrum change in the opposite 
sense \citep{muzerolle09,espaillat11}.  In rare cases this variability can be 
{\it mapped} in multi-epoch resolved scattered light observations 
\citep{stapelfeldt99,watson07a,wisniewski08}.  Models of the amplitudes, 
timescales, and wavelength dependence of the variability suggest an origin at 
sub-AU radii, where modulations of the vertical structure (due to quickly 
evolving non-axisymmetric features, turbulence, or warps) shadow the 
disk at larger radii \citep[e.g.,][]{flaherty10,flaherty13}.

\subsection{Radial Structure} \label{subsec:radial}

\subsubsection{Physical Overview}

The radial variation of the density structure is encapsulated in the {\it 
surface density} profile $\Sigma$ ($\equiv \int \rho~dz$).  As was already 
noted, $\Sigma$ is the most informative, essential aspect of disk structure.  
The shape of $\Sigma$ controls not only the likelihood of planet formation, but 
also the initial orbits and subsequent migrations of any planets during their 
formation epoch \citep[e.g.,][]{kokubo02,raymond05,miguel11}.  Moreover, 
$\Sigma$ offers an indirect snapshot of the mass and (angular) momentum flows 
that underpin the global structural evolution of disk material 
\citep[e.g.,][]{hartmann98}.  

The standard theoretical models for $\Sigma(r)$ are again based on the gas 
physics (presuming that the solids will follow the gas distribution; but see 
\hyperref[subsec:solidevol]{Sect.~\ref*{subsec:solidevol}}), using a balance of 
viscous and gravitational torques \citep{lin80,ruden86,ruden91,stepinski98}.  
For a thin Keplerian disk, $\Sigma$ (and its evolution) can be determined by 
solving the viscous diffusion equation \citep{lyndenbell74}
\begin{equation}
\frac{\partial \Sigma}{\partial t} = \frac{3}{r} \frac{\partial}{\partial r} 
\left[\sqrt{r} \frac{\partial}{\partial r} \left(\sqrt{r} \nu \Sigma \right)\right],
\label{eq:viscevol}
\end{equation}
where $\nu$ is the viscosity.  There is a vast literature associated with 
accretion flows and angular momentum transport in disks that centers around 
solving \cref{eq:viscevol} for a variety of viscosity distributions 
\citep[see the reviews by][]{pringle81,papaloizou95,turner14}.  The typical 
assumptions, at least in studies making connections to observational data 
\citep[although see][]{hueso05}, are that mass loss occurs only through 
accretion onto the star and that the viscosity is a (time-independent) radial 
power-law, $\nu \propto r^{\gamma}$.  The latter is motivated by the 
\citet{shakura73} prescription for turbulent viscosities in accretion disks, 
where $\nu \propto c_s^2/\Omega$.  In this case there are analytic similarity 
solutions to \cref{eq:viscevol}, with a $\Sigma$ profile that behaves like 
a power-law with an exponential taper at large radii,
\begin{equation}
\Sigma \propto \left(\frac{r}{r_c}\right)^{-\gamma} 
               \exp{\left[-\left(\frac{r}{r_c}\right)^{2-\gamma}\right]}.
\label{eq:simsoln}
\end{equation}
The characteristic scaling radius, $r_c$, and the normalization depend on time 
and the initial conditions \citep[e.g., size, mass, and mass flow 
rate;][]{lyndenbell74,lin87,hartmann98}.  

\begin{figure}[t!]
\epsscale{1.15}
\plotone{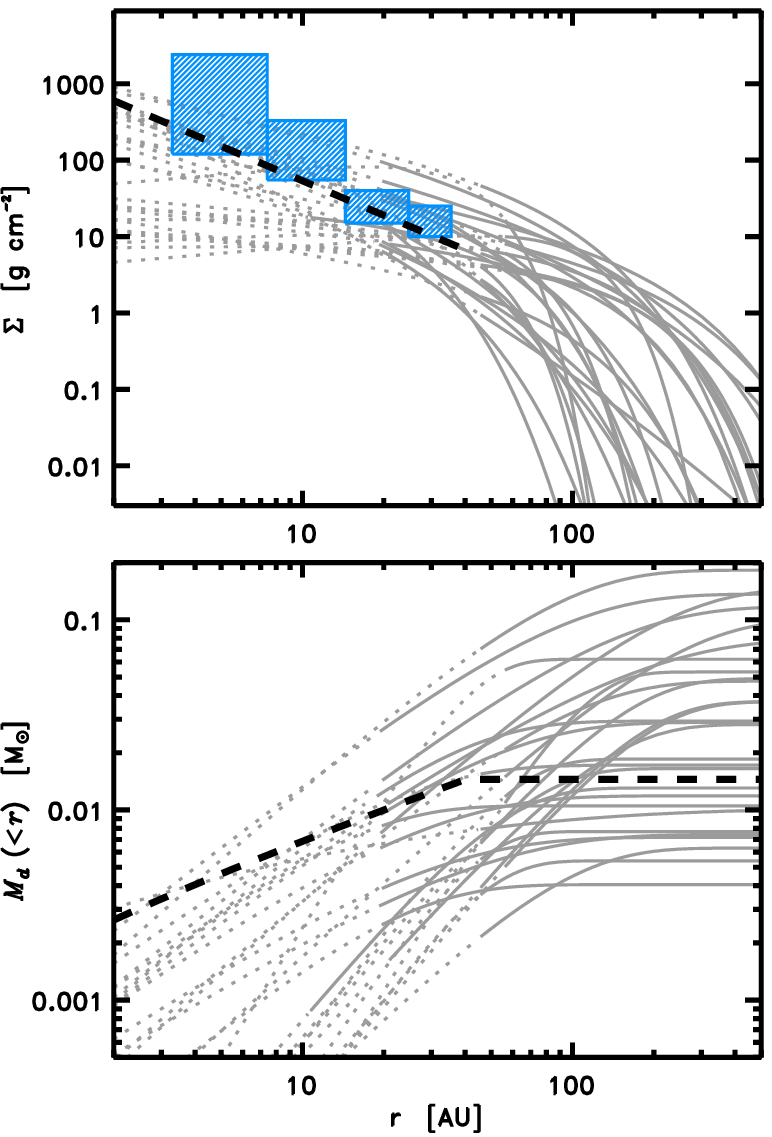}
\figcaption{({\it top}): The surface density profiles inferred from mm/radio 
interferometric observations of nearby disks \citep{andrews09,andrews10b,
isella09,isella10,guilloteau11}, assuming a spatially homogeneous opacity 
spectrum, a 1\%\ dust-to-gas ratio, and a similarity solution structure (cf., 
\hyperref[eq:simsoln]{Eq.~\ref*{eq:simsoln}}).  These profiles are broken at 
spatial scales that are not directly resolved (i.e., where $\Sigma$ has been 
extrapolated).  Blue boxes mark the standard \citet{weidenschilling77} 
construction of the MMSN around the giant planets; the dashed black curve shows 
the normalization and shape advocated by \citet{hayashi81}.  ({\it bottom}): 
The same information, but from the perspective of the cumulative mass 
distribution (the integrated mass inside some radius $r$).  
\label{fig:mmsn}}
\end{figure}

Given the uncertainty in these accretion disk models, empirical approximations 
are often preferred.  The most common is an old idea based on a simplistic 
reconstruction of $\Sigma$ in the solar nebula, where the current planet masses 
are augmented back to solar composition and then smeared into annuli 
\citep[e.g.,][]{young01,edgeworth49,kuiper56,cameron62,alfven70,kusaka70}.  For 
such a {\it minimum mass solar nebula} (MMSN) disk, $\Sigma$ is simply a 
truncated power-law: the canonical \citet{weidenschilling77} construction of 
the MMSN is shown in \Cref{fig:mmsn}.  Although there are fundamental 
conceptual flaws with the MMSN\footnote{Primary among these is the assumption 
of a static nebula and planetary system, meaning the MMSN density profile is a 
time-integrated structure (a convolution of two unknown effects: the density 
evolution and the efficiency by which material is accreted into the planets) 
rather than an instantaneous snapshot like would be observed in real disks or 
properly applied to a planet formation model \citep[see also][]{cameron88,
davis05,desch07}.}, it remains a standard benchmark in the field.  

\subsubsection{Observational Constraints}

The most direct constraints on surface density profiles come from resolved 
(interferometric) observations of the (sub-)millimeter continuum emitted by the 
disk solids (cf., \hyperref[subsec:thermal]{Sect.~\ref*{subsec:thermal}}).  
Such emission typically has low optical depths, meaning the surface brightness 
scales with the product $\kappa_{\nu} \, B_{\nu}(T) \, \Sigma$.  The standard 
approach to interpret the data is to assume that $\kappa_{\nu}$ is spatially 
homogeneous and then fit or solve for $T$ (as a radial power-law 
or with a radiative transfer calculation, respectively), and fit for $\Sigma$ 
(as in \hyperref[eq:simsoln]{Eq.~\ref*{eq:simsoln}} or a power-law), a 
characteristic or cutoff radius, and any other geometric parameters (e.g., 
inclination) with reference to the interferometric visibilities 
\citep[e.g.,][]{keene90,lay94,lay97,wilner96,wilner00b,wilner03,looney00}.  For 
marginally resolved data, there are strong degeneracies between the radius 
parameter, inclination, and the radial gradients in $T$ and $\Sigma$ 
\citep[e.g.,][]{mundy96}.  These can be mitigated with external constraints 
(the SED or scattered light data) on $T$ and/or geometry 
\citep[e.g.,][]{akeson02,kitamura02,pietu06,hamidouche06,aw07a,pinte08}, but 
they persist at problematic levels unless the brightness distribution is 
resolved well.  

For the nearest disk populations, a resolution of $\lesssim$0\farcs5 
(60--70\,AU) is usually sufficient to alleviate these degeneracies and place 
rudimentary constraints on $\Sigma$.  Surface density profiles have been 
inferred with such data for a few dozen disks, primarily in the Taurus 
\citep{isella09,isella10,pietu06,pietu14,guilloteau11} and Ophiuchus 
\citep{andrews09,andrews10b} regions (\cref{fig:mmsn}).  With access to radii 
$\gtrsim$15--30\,AU, these studies find a wide range of $\Sigma$ gradients 
($\gamma \approx 0$--1) and characteristic sizes ($r_c \approx 
5$--200\,AU)\footnote{There is an often unappreciated (although obvious) 
difference between cutoff radii in power-law models of $\Sigma$ and the 
characteristic radii ($r_c$) in the similarity solutions (cf., 
\hyperref[eq:simsoln]{Eq.~\ref*{eq:simsoln}}): the latter encircle some 
significant fraction of the mass, but there is a non-negligible extension of 
low-density material at larger radii.  Direct comparisons between different 
kinds of models require some care.}; some of that diversity reflects the 
different modeling approaches.  \Cref{fig:mmsn} also shows an alternative view 
in the form of cumulative mass distributions, which are perhaps more relevant 
for planet formation models.  Many disks have roughly sufficient masses at 
5--30\,AU (based on model extrapolations) to form planetary systems like our 
own.  However, the samples observed so far are biased (by necessity) toward the 
brightest, and thereby most massive, disks.

\begin{figure}[t!]
\epsscale{1.15}
\plotone{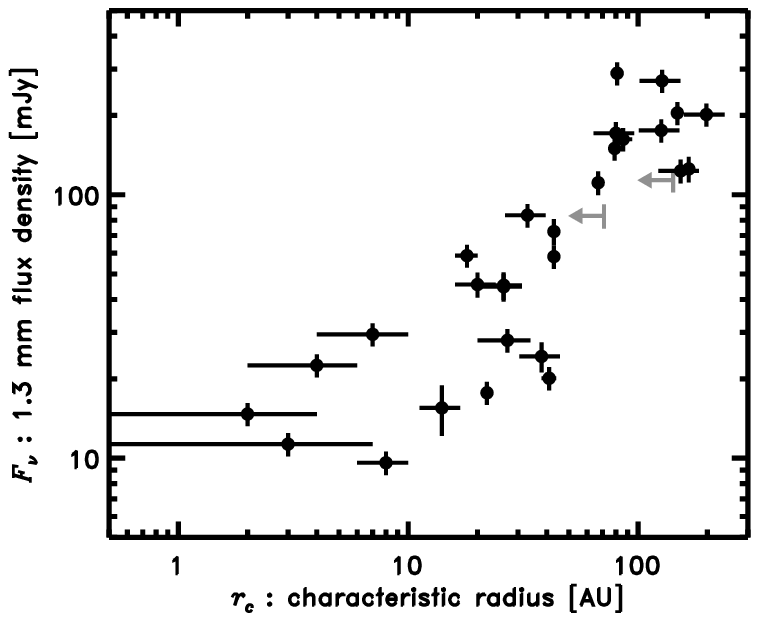}
\figcaption{The 1.3\,mm flux (scaled to a common distance), which scales with
$M_d$, as a function of the characteristic radius $r_c$ for all disks observed
with a resolution better than 0\farcs75 and modeled with the similarity
solution prescription for $\Sigma$
(\hyperref[eq:simsoln]{Eq.~\ref*{eq:simsoln}}).  Measurements are from
\citet{andrews09,andrews10b}, \citet{isella10}, \citet{guilloteau11}, and
\citet{pietu14}, and exclude binaries (see
\hyperref[subsec:multi]{Sect.~\ref*{subsec:multi}}) and transition disks (see
\hyperref[subsec:trans]{Sect.~\ref*{subsec:trans}}).
\label{fig:sizelmm}}
\end{figure}

\citet{andrews10b} noticed an interesting trend, now confirmed by 
\citet{pietu14}, that fainter disks are typically more compact; see 
\Cref{fig:sizelmm}.  This crude size--luminosity (mass) relationship could be a 
manifestation of initial conditions (diversity in core angular momenta) or 
viscous timescales \citep{andrews10b}; or perhaps it reflects 
variety in the evolutionary states of the solids (\citealt{pietu14}; see
\hyperref[subsec:solidevol]{Sect.~\ref*{subsec:solidevol}} for further 
discussion).  Following up on this trend is an active area of research, but 
larger samples and ancillary observations will be required to test hypotheses  
for its origins.

It is important to differentiate between what is being assumed and what is 
actually measured when interpreting resolved continuum data as described 
above.  The surface brightness scales roughly with $B_{\nu} 
(1-e^{-\tau_{\nu}})$ (\hyperref[eq:simpleRT]{Eq.~\ref*{eq:simpleRT}}), or $\sim 
B_{\nu}(T) \, \tau_{\nu}$ in the optically thin limit.  A constraint on 
$\Sigma$ requires a good (and internally self-consistent) model of the spatial 
variations of $T$ (i.e., that the vertical structure and irradiation source 
properties are well determined) and opacity (recall $\tau_{\nu} = \kappa_{\nu} 
\, \Sigma$).  For the latter, the typical assumption is that $\kappa_{\nu}$ is 
spatially invariant, which is {\it not} especially well-justified (see 
\hyperref[subsec:material]{Sect.~\ref*{subsec:material}}).  In fact, radial and 
vertical variations in $\kappa_{\nu}$ are generally expected 
(\hyperref[subsec:solidevol]{Sect.~\ref*{subsec:solidevol}}), and therefore 
will impact any estimates of $\Sigma$ and $T$.  While there are other caveats 
in this line of work, it is these basic systematic uncertainties about the 
material properties of the solids that ultimately limit the ability to robustly 
constrain disk $\Sigma$ profiles.

\subsection{Material Properties} \label{subsec:material}

It should be obvious at this point that the links between observations and the 
physical properties of disks are fundamentally tied to the detailed 
interactions of their constituent solid particles with radiation.  Those 
interactions are affected by the material properties of the solids, 
particularly their compositions, sizes, and morphologies.  This information is 
parametrically encoded in the ``optical" properties, absorption and scattering 
opacities as well as scattering and polarization phase functions, that control 
the redistribution of radiative energy in both frequency and direction.  These 
can be determined for any collection of particle properties and their 
corresponding optical constants (wavelength-dependent refractive indices) using 
numerical methods of varying sophistication, from simplified schemes like Mie 
theory \citep[e.g.,][]{bohren83} to the brute-force simulated solutions of 
Maxwell's equations in the discrete dipole approximation \citep{purcell73,
draine94}.  The optical constants for many relevant materials 
have been calculated theoretically \citep[e.g.,][]{draine84, laor93} or 
measured in the laboratory \citep[e.g.,][]{jaeger94,jaeger98,dorschner95,
henning95}.  

\subsubsection{Particle Composition}

Reasonable models for the composition of disk solids \citep[e.g.,][]{pollack94} 
are based on assumed chemical pathways and abundance measurements in primordial 
solar system bodies \citep{lodders03}: the primary contributors include 
silicates, carbonaceous materials, metallic compounds, and water ice.  The 
emission bands at $\sim$10 and 18\,$\mu$m in disk spectra verify an abundant 
population of small silicate particles with a range of mineralogies 
\citep{forrest04,kessler-silacci05,kessler-silacci06,watson09,sargent06,
sargent09,sargent09b,olofsson09,oliveira10,oliveira11}.  Additional features 
throughout the infrared provide evidence for hydrocarbons \citep[PAHs; 
e.g.,][]{habart04,geers06,geers07,keller08}, water ice \citep{chiang01,
creech-eakman02,mcclure15}, and other relevant minerals 
\citep[e.g.,][]{sturm13}.  Resolved maps of scattered light colors suggest 
organic grain coatings \citep[e.g., tholins;][]{debes08,debes13,rodigas15}.  
While it is not trivial to quantify the relative abundances of these materials, 
particularly in the midplane where spectral features are hidden by high optical 
depths, the combined constraints from disk spectra, analyses of meteorites, 
comets, and asteroids, and chemical intuition provide a reasonably coherent 
picture of the bulk composition of disk solids.  

\subsubsection{Particle Size Distribution}

The particle size (radius), $a$, dictates the cross-section for interaction 
with radiation of a given wavelength, and thereby has a significant impact on 
$\kappa_{\nu}$.  The opacity spectrum is approximately constant for $a \gg 
\lambda$ (geometric optics limit), falls off like $\lambda^{-2}$ when $a \ll 
\lambda$ (Rayleigh limit), and is enhanced by resonant interactions at 
intermediate wavelengths.  The implication is that emission observed at a given 
wavelength most efficiently probes particles of a similar size, $a \sim 
\lambda$.  In a realistic disk environment there will be a distribution of 
particle sizes, often approximated as a power-law, $dN/da \propto a^{-q}$.  
Typically, most of the mass is in the largest particles ($q < 4$); as a 
reference point, simple collisional models \citep{dohnanyi69} and fits to the 
interstellar extinction curve \citep{mathis77,cardelli89,mathis90} suggest $q 
\approx 3.5$.

\begin{figure}[t!]
\epsscale{1.2}
\plotone{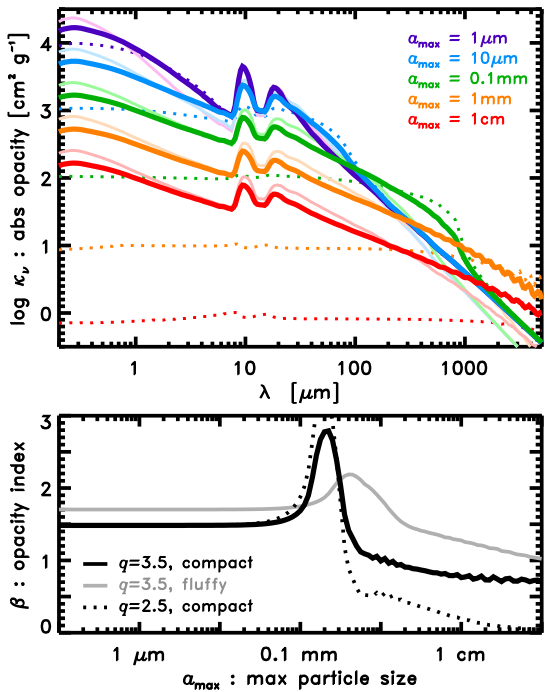}
\figcaption{({\it top}): Examples of absorption opacity spectra for simple 
models of dust populations relevant for disks, based on the homogenized 
analysis efforts of the {\tt DIANA} project \citep{woitke15}.  The composition 
is a mixture of amorphous silicates, carbon, and vacuum, with various 
assumptions about the size distribution and porosity (in this case, ``compact" 
and ``fluffy" mean filling factors of 1 and 0.5, respectively).  Optical 
constants were collated from \citet{dorschner95} and \citet{zubko96}, and 
combined using the \citet{bruggeman35} mixing rule.  The `distribution of 
hollow spheres' approach advocated by \citet{min05} was then used to compute 
opacities.  ({\it bottom}): The 1--3\,mm opacity spectral index, $\beta$, as a 
function of the maximum particle size.  The typical spectral slopes measured 
for disks cluster around $\beta \lesssim 1$, indicating that most of the 
particle mass has grown to at least mm sizes.
\label{fig:opacities}}
\end{figure}

\Cref{fig:opacities} illustrates how the opacity spectrum changes for different 
particle size distributions.  As the maximum size increases (or the 
distribution becomes more top-heavy), there are three general changes to the 
opacities.  First, the infrared opacities decrease.  However, if the disk is 
still optically thick in the infrared, it is not easy to demonstrate 
this observationally.  Second, the silicate spectral features become muted 
\citep[e.g.,][]{min04}.  This has been identified as a signature of grain 
growth in the disk surface layers \citep[e.g.,][]{bouwman01,meeus01,
kessler-silacci06,olofsson09,lommen10}, although there is room for 
structural \citep[e.g.,][]{dominik08} or compositional \citep{honda03,meeus03} 
ambiguity in quantifying the size distribution.  And third, the approximately 
power-law spectral behavior at mm/radio wavelengths flattens; i.e., if 
$\kappa_{\rm mm} \propto \nu^{\beta}$, then $\beta$ decreases with larger 
$a_{\rm max}$ or smaller $q$ (see \cref{fig:opacities}; \citealt{miyake93,
ossenkopf94,henning96,dalessio01,draine06}).  

This latter point is especially interesting for disks because they become 
optically thin at these long wavelengths.  The spectral dependence of the 
mm/radio continuum therefore scales with $B_{\nu} \, \kappa_{\nu}$; in the 
Rayleigh-Jeans limit the observed spectrum would scale like $\nu^{2+\beta}$, 
and so the radio ``colors" serve as a key diagnostic of the particle size 
distribution.\footnote{In practice, the Rayleigh-Jeans limit is often not 
strictly valid for disks.  However, even simple assumptions for $T$ can be used 
to appropriately account for the curvature in the Planck term.}  For the small 
(sub-micron) grains in the interstellar medium, the mm/radio spectrum is steep; 
$\beta \approx 1.8\pm0.2$ \citep{hildebrand83,goldsmith97,finkbeiner99,li01}.  
The integrated spectra of $\sim$Myr-old disks have comparatively much ``redder" 
colors, indicative of $\beta \lesssim 1$ in most cases \citep{beckwith91,
mannings94,aw05,wilner05,lommen07,lommen09,ricci10a,ricci10b,ubach12} and 
therefore substantial growth \citep[$a_{\rm max}$$\sim$mm or cm; 
e.g.,][]{draine06}.  There are a few important points to keep in mind regarding 
the interpretation of spectral behavior in this context.  For shallow spectra 
(low $\beta$), the relationship between the opacity spectral index and particle 
size saturates (see \cref{fig:opacities}); when solids reach sizes much larger 
than the observing wavelengths, they emit much less and have negligible impact 
on the observed spectrum.  So, associating $\beta$ with a particle size 
distribution extending beyond cm-sized solids is generally not possible.  More 
practically, it is crucial to spatially resolve brighter disks, so any 
contamination from optically thick emission (with a spectrum that scales like 
$\nu^2$) can be identified \citep{testi03,natta04,rodmann06,ricci12}.  
Likewise, when increasing the spectral range to the radio ($\lambda \sim 
1$\,cm) in an effort to better leverage constraints on $\beta$, the 
contamination of {\it non-disk} emission (free-free or synchrotron radiation) 
also needs to be considered \citep[e.g.,][]{rodmann06}.  

\subsubsection{Particle Morphology}

The optical properties of solids are also sensitive to their structure or 
morphology, typically discussed in terms of porosity \citep[or the volume 
filling factor; e.g.,][]{kozasa92,henning96,min03,kimura03,ormel11}.  Numerical 
simulations predict that small individual grains (monomers) will stick together 
in low-speed collisions and grow into fluffy, porous aggregates 
\citep[e.g.,][]{ossenkopf93,dominik97,ormel07,wada09,okuzumi12}; laboratory 
experiments directly confirm this behavior \citep[see][]{blum08}.  The optical 
properties of porous aggregates depend on the product of their bulk size and 
filling factor \citep[e.g.,][]{kataoka14}.  The absorption opacities of large 
aggregates with high porosity are similar to those for smaller, compact 
particles, with the subtle difference being that the resonant enhancement of 
$\kappa_{\nu}$ for $a \sim \lambda$ is muted (cf., \cref{fig:opacities}).  The 
corresponding scattering opacities may be relatively enhanced at long 
wavelengths: \citet{kataoka14} suggest mm-wave polarimetry as a potential probe 
\citep[although current limits are already quite strong; e.g.,][]{hughes09b,
hughes13}.  More information on the morphologies of disk solids are available, 
at least in the surface layers, through constraints on the scattering phase 
function \citep[e.g.,][]{mulders13} and polarized intensities 
\citep[e.g.,][]{min12}.  

\subsubsection{Caveats on Mass and Density Estimates}

On consideration of all these complex microphysical factors that set the 
opacities, which in turn are responsible for translating physical conditions 
into the astronomical tracers that are observed, the effort to quantify the 
role of disks in planet formation may seem incredibly daunting.  And indeed, 
there is much work to be done in teasing out the complex inter-dependences 
between the material and structural properties of disks.\footnote{The situation 
is in fact more complicated than presented, since the compositions, sizes, and 
morphologies of the solids all vary as a function of the physical conditions 
(e.g., see \hyperref[subsec:solidevol]{Sect.~\ref{subsec:solidevol}}).}  So, it 
is imperative to consider inferences about disk properties in the context of 
the {\it assumed} (i.e., model-dependent) properties of their constituent 
solids.  

Given the fundamental significance of disk masses in the planet formation 
process, there is special interest in understanding the uncertainty (or 
diversity) in the opacities at millimeter wavelengths.  With the few exceptions 
of disks with especially simple structures \citep[e.g.,][]{andrews14}, this 
issue is best assessed demographically.  A few different arguments have been 
made that point toward the ``standard" millimeter opacities \citep[$\sim$2 
cm$^2$\,g$^{-1}$ at 1.3\,mm;][]{beckwith90} being over-estimated, and therefore 
that the disk masses may be higher than are typically inferred.  One points out 
the order of magnitude mismatch between the masses inferred from the millimeter 
emission and a crude proxy using the product of accretion rates and ages 
\citep{hartmann98,hartmann06}.  Another derives from the claim that the massive 
disk fraction seems to be too low to produce the observed giant planet 
frequency \citep[e.g.,][]{greaves10,najita14}.\footnote{This discrepancy is not 
too large, since roughly a quarter of $\sim$Myr-old disks around Sun-like hosts 
have masses in excess of the MMSN (\citealt{andrews13}; \cref{fig:mdms}).  
Nevertheless, it demonstrates that the masses are likely not systematically 
over-estimated.}  The {\it direction} of this proposed uncertainty makes sense, 
since the growth of solids will lock up a fraction of the mass in large bodies 
that do not emit much at millimeter wavelengths (cf., \cref{fig:opacities}).  
But its {\it scale} is perhaps less clear.  It is unlikely that the masses 
could be substantially higher than estimated without a large population of 
gravitationally unstable disks, for which no evidence 
\citep[e.g.,][]{narayanan06,jang-condell07,dipierro14} has yet been identified 
\citep[see especially the symmetry for the very bright HL Tau 
disk;][]{brogan15}.  So considering these (admittedly approximate) constraints, 
it seems reasonable to assume that the systematic uncertainty on the mean 
millimeter opacity in disks (relative to the typically adopted value) is 
modest, about ($\pm$)1\,dex.

\subsection{Synopsis}

\begin{itemize}

\item The flared, vertical distribution of dust grains regulates the thermal 
structure of a disk.  The broadband infrared SED and spatially resolved 
measurements of optical/infrared scattered light can be used to constrain the 
vertical structure.  In the context of simple models, such data suggest that 
the typical vertical aspect ratio ($\sim$$H/r$) is 0.1.

\item The radial surface density profile, $\Sigma$, is a crucial input for 
planet formation models.  Interferometric measurements that resolve the 
optically thin millimeter continuum emission can be used to interpret the 
radial distribution of solid mass; some preliminary results are consistent with 
models of the (outer) primordial solar nebula.  

\item The compositions, sizes, and shapes of the solid particles determine the 
opacities, which regulate the radiative transfer of energy into and out of the 
disk structure.  There are strong and complex inter-dependencies between these 
material and structural properties.  Measurements of the shape of the mm/radio 
continuum spectrum are sensitive to the particle size distribution; the 
relatively ``red" mm/radio colors that are observed indicate the substantial 
growth of disk solids.

\item The physical conditions inferred from the interpretation of disk 
observations need to be considered only in the context of the model assumptions 
(e.g., the functional form of the density distribution, the material properties 
of the constituent solids).

\end{itemize}

\noindent {\it Additional Reading}: the review of disk modeling by 
\citet{dullemond07}; a basic review of resolved structure constraints by 
\citet{wilner00}; the more general review by \citet{williams11}.

\section{Evolution} \label{sec:evolution}

The general framework for interpreting protoplanetary disk observations, as 
outlined in the previous sections, is still rather crude and over-simplified.  
With some recent enhancements in data quality, it has become clear that a more 
nuanced approach is necessary.  There is now unequivocal evidence that 
evolutionary effects play a large (and likely dominant) role in determining the 
observable characteristics of these disks.  The goal of this section is to 
highlight how two key evolutionary behaviors -- the growth and migration of 
solids, and the relatively rapid clearing of the inner disk -- impact the 
standard observational benchmarks of disk structure, and thereby to point out 
opportunities for new insights on the physical mechanisms that drive such 
evolution.

\subsection{The Growth and Migration of Solids} \label{subsec:solidevol}

Most observational studies of disk structures implicitly assume that the 
collisional evolution of the constituent solids can be ignored.  That is a bad 
assumption.  The growth and migration of these solids involve extraordinarily 
complex processes, but exploring their consequences -- both physically and 
observationally -- is a crucial step toward developing a more comprehensive 
understanding of planet formation.  To illustrate the effects of this 
evolution, in this section we will consider what happens to an `initial' 
population of small dust grains suspended in a (hydrostatically supported) gas 
reservoir with a homogeneous dust-to-gas ratio, $\zeta(r, z) = \zeta_0$.  

\subsubsection{Settling}

Relative motions imparted stochastically (Brownian motion, turbulent mixing) 
will make these grains collide; at low speeds, they will stick together and 
produce larger aggregates \citep[e.g.,][]{blum08,guttler10}.  Smaller particles 
are coupled to the gas, buoyed up into the disk atmosphere by hydrostatic 
support.  But larger particles feel less of that lift and will settle toward 
the midplane \citep{adachi76,weidenschilling77,nakagawa81}.  As they sink into 
denser regions, the collision rate increases and growth can accelerate.  This 
growth and (vertical) migration feedback process is not perfectly efficient 
\citep{dullemond05}: alternative collision outcomes like electrostatic 
repulsion \citep{okuzumi09} and bouncing \citep{zsom10} can slow growth, while 
fragmentation \citep{brauer08,birnstiel09} and erosion 
\citep[e.g.,][]{seizinger13,krijt15} can actually reverse it.

The interactions between the gas pressure, gravity, and (turbulent) mixing, 
coupled with the efficiency of particle growth (relative to destruction), gives 
rise to a size-sorted, sedimentary layering of the disk solids in which $\zeta$ 
{\it decreases} with $z$ \citep[e.g.,][]{volk80,cuzzi93,dubrulle95,
schrapler04}.  This size segregation increases the vertical temperature 
gradient and concentrates the absorption/scattering optical depth profile 
toward the midplane, reducing the irradiated 
surface area and height of the surface layer and thereby impacting some key 
observables (e.g., \citealt{dullemond04b}; see 
\hyperref[subsec:vertical]{Sect.~\ref{subsec:vertical}}).

A {\it settled} disk has a lower characteristic surface layer, absorbs less 
starlight, and therefore is colder.  Consequently, it emits less thermal and 
scattered continuum radiation to the observer \citep{miyake95,dullemond04b,
dalessio06,mulders12}.  Distinguishing settling from a disk with an 
intrinsically low gas pressure scale height, $H$, is not trivial.  Some 
indirect constraints can be made from comparisons of mid-infrared colors and 
the strength of the 10\,$\mu$m silicate feature 
\citep[e.g.,][]{kessler-silacci06,furlan09}, or the inconsistent inferences of 
surface height layers from different tracers \citep[e.g., scattered light 
images and infrared SEDs;][]{stapelfeldt03,wolf03}.  A more direct signature of 
this stratification is accessible by comparing scattered light images at 
multiple wavelengths.  Since dust grains preferentially scatter at wavelengths 
comparable to the particle size, settled disks show a scattering layer height 
that decreases with $\lambda$ \citep{duchene03,duchene04,duchene10,pinte07,
grafe13}.  Edge-on disks may be particularly useful in future work on this 
subject, when ALMA can spatially resolve both the temperature structure of 
the gas \citep[e.g.,][]{dartois03,rosenfeld13a,degregorio-monsalvo13} and the 
continuum that traces $\sim$mm particles as a function of 
height above the midplane \citep[e.g.,][]{boehler13}.  As a reference, 
radiative transfer models of typical disk SEDs indicate that settling can 
reduce the dust-to-gas ratio in the disk atmosphere by a factor $1/\epsilon 
\approx 10$--1000, where $\zeta = \epsilon \, \zeta_0$ in the parlance of 
\citet{dalessio06}.  

\begin{figure}[t!]
\epsscale{1.15}
\plotone{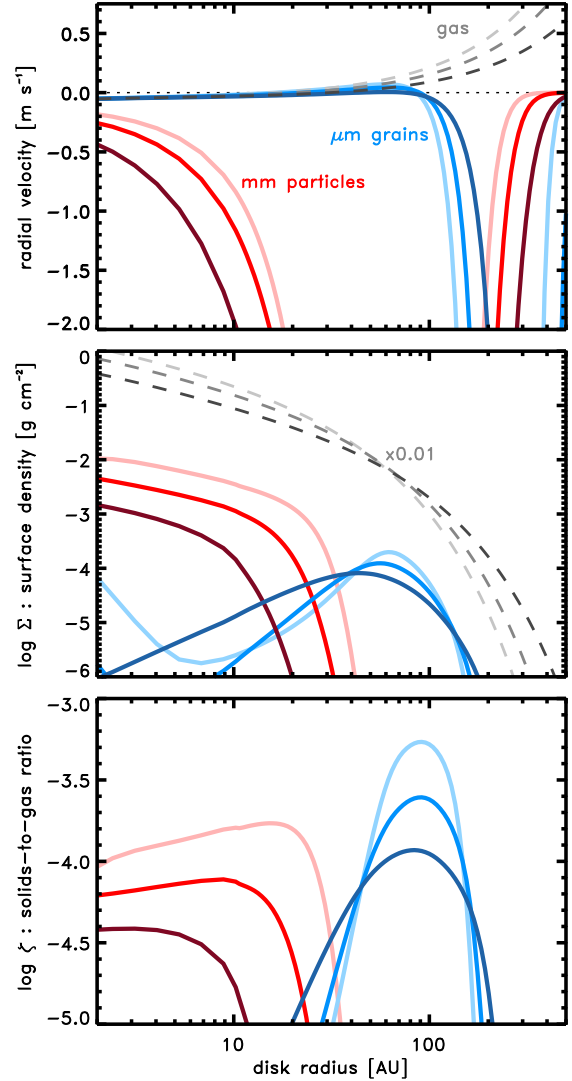}
\figcaption{Demonstration of the evolution of solids embedded in a gas-rich 
disk.  In each panel, grays refer to the gas, blues to $\sim$$\mu$m-sized 
grains, and reds to $\sim$mm-sized particles.  Three time steps are shown -- 
0.5, 1, and 2\,Myr (from light to dark) -- starting from an initial homogeneous
distribution of sub-$\mu$m monomers with $\zeta = 0.01$ in a standard viscous 
accretion disk (see \citealt{birnstiel14} for details).  ({\it top}) Radial 
velocities: large negative (inward) motions (up to $\sim$0.1\,km s$^{-1}$ 
levels) 
due to radial drift are noted for mm-sized solids at tens of AU.  ({\it 
middle}) Surface densities ($\Sigma$): the mm solids are radially concentrated 
and have a relatively sharp ``edge" well interior to the gas or small grain 
distributions.  ({\it bottom}) Dust-to-gas mass ratio ($\zeta$): a steep drop 
in $\zeta_{\rm mm}$ is produced by radial drift.  But note the scale in 
$\zeta$: these low values are emblematic of a major issue with the radial drift 
timescales (they are much too short) in current evolutionary models.
\label{fig:drift}}
\end{figure}

\subsubsection{Radial Drift}

Similar evolution also proceeds in the radial dimension.  The radial motions of 
disk solids are fundamental to that evolution.  Particles that are well coupled 
to the gas will follow its viscous flow trajectories, moving inward toward the 
host star at small radii (accretion), and away from it in the outer disk 
(diffusion) \citep[e.g., see][]{takeuchi02}.  These radial velocities induced 
by viscous drag tend to be slow ($\sim$few to tens of cm s$^{-1}$), and are 
most relevant for small ($\sim$$\mu$m-sized) particles.  An aerodynamic process 
termed ``radial drift" can be substantially more influential.  In general, the 
gas motions in a disk are dominated by rotation around the host star.  However, 
the presence of a radial pressure gradient imparts a small deviation from 
Keplerian velocities.  For example, in the standard picture of a smooth disk 
with temperatures and densities that decrease with $r$, the pressure gradient 
is negative and the gas disk rotates at a slightly sub-Keplerian rate.  In the 
limit of weak coupling to the gas, solids orbit at Keplerian speeds.  In the 
intermediate drag regime, particles with a certain range of sizes feel a strong 
headwind due to their velocity differential with the gas, which saps their 
orbital energy and sends them spiraling toward the pressure maximum 
\citep{whipple72,adachi76,weidenschilling77,nakagawa81,nakagawa86}.  This 
radial drift moves solids so efficiently that they can be locally depleted 
faster than the coagulation timescale: in a MMSN disk, such a `drift barrier' 
to further growth occurs at $\sim$1\,AU for m-sized bodies, and at 
$\sim$100\,AU for mm-sized particles.

The radial migration outlined above significantly modifies the distribution of 
disk solids in both size and space.  \Cref{fig:drift} illustrates these effects 
for a representative evolutionary model \citep[cf.,][]{birnstiel14}.  The 
higher densities and relative velocities present at smaller disk radii 
naturally enhance the local collision (and thereby growth) rates 
\citep[e.g.,][]{dullemond05}.  But the added influx of $\sim$mm/cm-sized 
particles from the outer disk will amplify this radial size-sorting.  The net 
result is that larger solids will be preferentially concentrated close to the 
gas pressure maximum \citep[typically the inner disk edge; e.g.,][]{takeuchi02,
brauer08,birnstiel10,birnstiel12}.  Since most of the solid mass is in larger 
particles that are systematically depleted from the outer disk by radial drift, 
the dust-to-gas ratio $\zeta$ decreases with $r$ \citep{takeuchi05b,
birnstiel14}.  For the sizes where drift velocities are maximal in the outer 
disk ($\sim$mm/cm), the ratio ($\zeta_{\rm mm}$) drops precipitously at 
$\sim$10--100\,AU \citep{youdin02,weidenschilling03,takeuchi05,jacquet12,
birnstiel14}.  

\subsubsection{Signatures of Radial Drift}

\begin{figure}[t!]
\epsscale{1.15}
\plotone{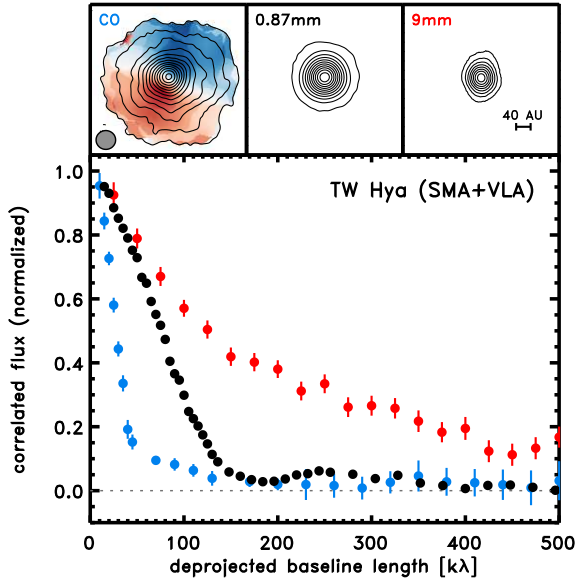}
\figcaption{Illustration of the observational effects of particle growth and 
migration in the TW Hya disk.  ({\it top, left to right}) The CO $J$=3$-$2 
line (color scale is the intensity-weighted velocity field), 870\,$\mu$m 
continuum, and 9\,mm continuum emission observed by the SMA \citep{andrews12} 
and VLA \citep{menu14} interferometers.  The images have the same scale 
(4\arcsec, $\sim$430\,AU, on a side) and resolution (1\arcsec, $\sim$50\,AU).  
Note the progression of radial concentration in the various tracers that is 
predicted by theoretical models (cf., \cref{fig:drift}).  ({\it bottom}) The 
same progression from extended (CO) to compact (9\,mm continuum), now viewed 
more directly as the azimuthally averaged visibility profiles (see 
\hyperref[subsec:thermal]{Sect.~\ref*{subsec:thermal}} and 
\cref{fig:vis_demo}).  The low-level oscillations in the 870\,$\mu$m 
visibilities are a distinct signature of the ``sharp" edge in the radial 
distribution of $\sim$mm-sized particles imposed by radial drift (see 
\cref{fig:drift}, {\it bottom}, and \citealt{birnstiel14}).  
\label{fig:obs_drift}}
\end{figure}

These physical effects have pronounced observational hallmarks.  The radial 
sorting of particle sizes induced by growth and drift is manifested in the disk
opacities, $\kappa_{\nu}$, particularly at mm/cm wavelengths (cf., 
\hyperref[subsec:material]{Sect.~\ref*{subsec:material}}).  Since larger 
particles have a lower opacity spectral index, $\beta$ (a ``redder" 
$\kappa_{\nu}$), disks that have experienced substantial evolution in their 
solids will have an {\it increasing} $\beta(r)$: such an opacity profile is 
observed as a radial variation of the {\it radio colors} \citep[i.e., a 
relatively ``blue", or steep, spectrum at larger $r$;][]{isella10}.  In 
practice, this color gradient is observable as an anti-correlation between the 
wavelength and size of the continuum emission, such that the intensity
distribution at longer wavelengths is more compact \citep{banzatti11,
guilloteau11,perez12,lperez15,trotta13,menu14}.  \Cref{fig:obs_drift} shows an 
example of this effect.

A confirmation of the related prediction that drift produces a decreasing 
$\zeta(r)$ (cf., \cref{fig:drift}) is more difficult, since accessing and 
interpreting gas density tracers remains a challenge.  However, some indirect, 
qualitative evidence supporting a steep drop-off in $\zeta_{\rm mm}(r)$ is 
available through a comparison of the mm/cm-wavelength continuum and molecular 
line emission {\it sizes}: the former should have a more radially compact 
distribution than the latter.  Such line/continuum size discrepancies have been 
noted for some time \citep[e.g.,][]{pietu05,isella07}, but were (reasonably) 
attributed to optical depth effects: since the mm/cm continuum is optically 
thin, insufficient sensitivity can lead to a size under-estimate relative to 
the optically thick lines that typically trace the gas \citep{hughes08}.  As 
the data quality has improved, it has nevertheless become clear that these 
apparent size discrepancies are not observational artifacts, but are instead 
intrinsic to the disk structures.  For a still-limited sample, this signature 
of radial drift is manifested in a mm continuum distribution $\sim$2--5$\times$ 
smaller than for the CO line (\citealt{panic09,andrews12,rosenfeld13b,
degregorio-monsalvo13,walsh14}; see \cref{fig:obs_drift}).

Some complementary information is encoded in the morphology of the mm continuum 
at larger radii.  The steep drop in $\zeta_{\rm mm}(r)$ predicted theoretically 
to be a consequence of radial drift (e.g., \citealt{birnstiel14}; see 
\cref{fig:drift}) should produce a relatively sharp outer ``edge" in the 
continuum distribution.  That edge is clearly identifiable in the oscillatory 
structure it imprints on the interferometric visibilities (see 
\cref{fig:obs_drift}; \citealt{andrews12,degregorio-monsalvo13}).  

\subsubsection{Open Issues}

The dramatic modifications these evolutionary factors impart on the traditional 
observational tracers of structure (cf., 
\hyperref[sec:structure]{Sect.~\ref*{sec:structure}}), should make it clear 
that constraining disk densities is a substantially more complicated task than 
has often been assumed.  At some level, the spatial variations imposed on 
$\kappa_{\nu}$ and $\zeta$ by these evolutionary mechanisms render many of the 
standard assumptions used in the interpretation of disk structures invalid.  To 
put it bluntly, there is not currently any obvious, coherent parametric 
framework available for characterizing the density distributions of disk 
solids.  Right now, the focus is instead on collecting additional evidence that 
supports the evolution of solids, and building up a foundation for its proper 
interpretation.  The expectation is that resolved multiwavelength continuum and 
spectral line data can be combined to better characterize the growth and 
migration of solids, and thereby to facilitate integrated studies of structure 
and evolution.

This subject is a relatively new, and incredibly dynamic, emphasis for the 
field.  But while the influx of high-quality resolved data is {\it 
qualitatively} in sync with theoretical predictions for the growth and 
migration of disk solids, it is important to recognize that there are also some 
serious conflicts.  Most prominent among these is the issue of migration 
timescales.  Theory predicts that radial drift is {\it much} too efficient to 
explain the spatial distributions and luminosities of the mm-wavelength 
continuum that are routinely observed for $\sim$Myr-old disks 
\citep[e.g.,][]{takeuchi05,brauer07}.  The hypothesized remedy to relieve this 
tension is that there must be small-scale substructure in the gas disk 
\citep[e.g.,][]{whipple72,pinilla12a}.  Localized pressure maxima introduced by 
turbulent fluctuations \citep{klahr97,dzyurkevich13,flock15}, 
magnetohydrodynamic (MHD) structures (i.e., zonal flows; \citealt{johansen09,
bai14b,suzuki14,bai15}), condensation fronts \citep[e.g.,][]{kretke07}, or 
dynamical perturbations from (planetary) companions \citep{pinilla12b,
birnstiel13} can slow or even ``trap" drifting particles, potentially 
reconciling the theoretical timescales and observational constraints.  There is 
some preliminary evidence in support of these kinds of substructure for a 
special subset of disks (see 
\hyperref[subsec:trans]{Sect.~\ref*{subsec:trans}}), and there will soon be 
data capable of quantifying its nature and prevalence in the more general disk 
population (see \hyperref[sec:future]{Sect.~\ref*{sec:future}}).

\subsection{The ``Transition" Phase} \label{subsec:trans}

\subsubsection{Physical Overview}

The evolution of angular momentum, through turbulent transport 
\citep{shakura73,lyndenbell74,hartmann98} or dissipation via magnetic winds 
\citep{blandford82,konigl89,bai13}, largely dictates the long-term behavior of 
disk structures.  If acting alone, such processes would steadily deplete the 
densities until the disk slowly (tens of Myr) faded away.  However, there are 
additional mechanisms that should expedite the dissipation or transformation of 
the disk material.

In terms of dispersal, the primary pathway of interest is through a 
photoevaporative wind.  Eneregtic radiation (far-ultraviolet to X-ray) 
generated by the host star can impart the gas in the disk surface with enough 
energy to escape the system \citep{clarke01,alexander06a,ercolano08,gorti09,
owen10}.  When the mass-loss rate in the resulting wind exceeds the (inward) 
viscous flow rate, a gap opens near the wind launch radius \citep[$\sim$few AU; 
e.g.,][]{liffman03,font04,ercolano09}.  The inner disk is cut off from 
re-supply, and will quickly ($\lesssim$0.1\,Myr) accrete onto the star 
\citep[$\sim$0.1\,Myr; e.g.,][]{alexander06b,gorti09b,owen11}.  This depletion 
makes the inner disk optically thin in the radial direction, permitting a more 
efficient irradiation at the inner edge of the remaining disk.  This in turn 
enhances the wind mass-loss rate and thereby quickly ($\lesssim$0.01\,Myr) 
expands the wind-driven ``cavity" at the disk center to $r\gtrsim10$\,AU 
\citep{clarke01,alexander06b,owen12}; it may even trigger an instability that 
destroys the disk altogether \citep{owen13}.  

In the alternative \citep[but not mutually exclusive; e.g.,][]{matsuyama03,
rosotti13} context of metamorphosis, planet formation can also dramatically 
alter the disk structure on relatively short timescales.  Once a planet is 
sufficiently massive (depending on the disk viscosity, but typically 
$\sim$0.1--1\,M$_{\rm jup}$), it exerts torques that repel the local disk 
material away from its orbital radius \citep[e.g.,][]{goldreich80,lin79,lin93,
bryden99,kley12}.  The net result is a narrow annulus of depleted densities, a 
gap.  Depending on the local disk properties and planet mass, some disk 
material may continue to flow to the planet (or beyond it) in accretion streams 
\citep{artymowicz96,kley99,lubow99}.  A more massive planet accretes a larger 
fraction of this flow \citep[e.g.,][]{lubow06}, diminishing the re-supply of 
the disk region interior to the gap, and thereby effectively creating a 
``cavity" at the disk center \citep[e.g.,][]{varniere06,crida07,
dodson-robinson11,zhu11}.  

Theoretical models of these processes make some important and testable 
observational predictions.  The most fundamental of these is that disk 
evolution should involve two timescales --- a long stage where the processes 
related to momentum transport dominate, followed by a rapid ``transition" phase 
where the disk structure is severely modified by interactions with winds and/or 
(giant) planets --- that can be identifiable in simple demographic data 
\citep[e.g., see][]{alexander09}.  The disks in this latter evolutionary phase 
have a distinct structural feature: a central region of depleted densities.  
Such a structure necessarily imposes a pressure maximum just outside this 
low-density cavity, which may concentrate drifting solids in a ring-shaped 
``trap" (cf., \hyperref[subsec:solidevol]{Sect.~\ref*{subsec:solidevol}}; e.g., 
\citealt{alexander07,pinilla12b,gorti15}).  As with normal disks (cf., 
\hyperref[subsec:solidevol]{Sect.~\ref*{subsec:solidevol}}), these solids leave 
behind a radially extended gas disk with a low dust-to-gas ratio.  The shape 
of the pressure gradient around the maximum regulates the flow of material to 
smaller radii.  If the gradient is not too steep, gas and smaller particles can 
pass through into the cavity; however, larger solids are filtered out of this 
flow and remain trapped near the pressure maximum \citep[e.g.,][]{rice06,
paardekooper06,zhu12,zhu14}.  In a sense, the small amount of material in the 
cavity (or, rather, the flow rate of disk material through that pressure 
maximum), provides important clues to the physical mechanism(s) responsible for 
the depletion that is central to this transitional phase of evolution
\citep[e.g.,][]{alexander06b,najita07,najita15,alexander09,owen12b}.  

\begin{figure}[t!]
\epsscale{1.15}
\plotone{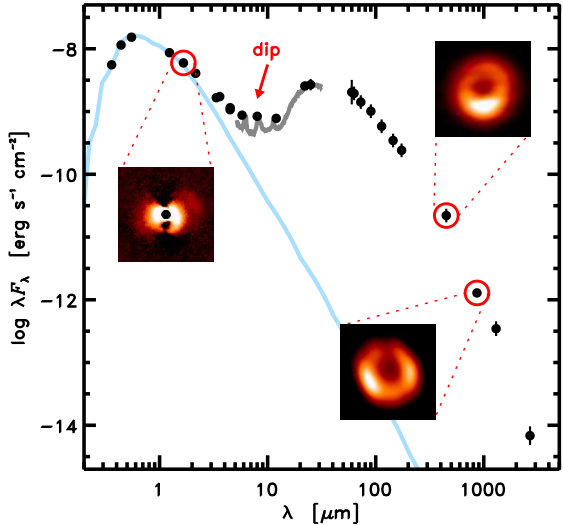}
\figcaption{The SED and representative resolved images of the transition disk 
around the young star SR21.  The blue curve is a model of the stellar 
photosphere; the gray curve is a well-sampled spectrum from {\it Spitzer}.  The 
SED shows a prominent ``dip" in the infrared, produced by the substantial 
depletion of a cavity with radius $\sim$20\,AU.  The image insets show the 
1.6\,$\mu$m (polarized) scattered light \citep{follette13}, along with the 
450\,$\mu$m \citep{perez14} and 870\,$\mu$m thermal continuum emission 
\citep{brown09}.  Each image spans 175\,AU on a side.  The thermal images 
clearly show the low-density cavity, but there is scattered light tracing small 
grains located well inside the (sub-)mm ring.
\label{fig:trans_rings}}
\end{figure}

\subsubsection{Observational Signatures}

The observational evidence for this rapid stage of disk evolution pre-dates 
(and motivates) much of the theoretical work.  In any young cluster, there is a 
small sub-population ($\sim$1--10\%) that lacks a near-infrared excess but 
still exhibits strong dust emission at longer wavelengths \citep{strom89,
skrutskie90}.  These {\it transition disks} are usually identified by the 
``dip" in their infrared spectra, with blue colors shortward of 
$\sim$5--10\,$\mu$m and red colors at longer wavelengths 
\citep[e.g.,][]{brown07,cieza07,cieza12,merin10,furlan11}.  A representative 
example is shown in \Cref{fig:trans_rings}.  Given the rough $\lambda \mapsto 
r$ mapping between the spectrum and dust temperatures 
(\hyperref[subsec:thermal]{Sect.~\ref*{subsec:thermal}}), this ``dip" signature 
suggests a substantial depletion of solids in the warm inner disk 
\citep{strom89,skrutskie90,marsh92,marsh93,calvet02,calvet05,rice03,
dalessio05}.  If all disks experience such a transition, a simple duty cycle 
argument demands that it occurs quickly (the product of the transition disk 
fraction and the mean sample age, so $\lesssim$0.1--0.5\,Myr; but see 
\citealt{sicilia-aguilar08,muzerolle10}).  Comparing that rough timescale with 
the decay time of infrared excesses in the general disk population 
(\hyperref[subsec:massevol]{Sect.~\ref*{subsec:massevol}}; see 
\citealt{mamajek09}), these simple infrared color demographics clearly reveal 
the theoretically predicted two-timescale behavior for disk evolution.

\begin{figure}[t!]
\begin{center}$
\begin{array}{cc}
\includegraphics[width=1.55in]{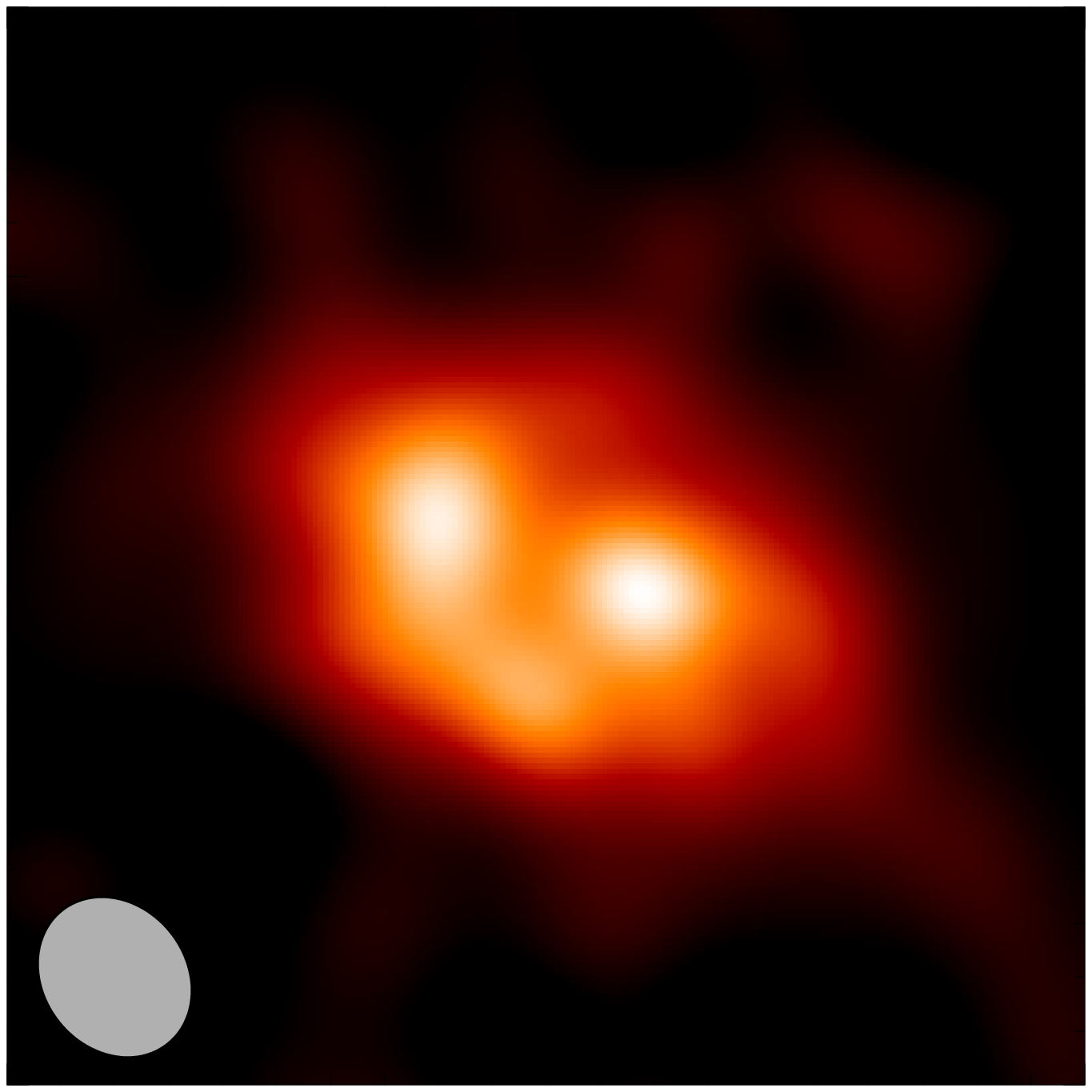} &
\includegraphics[width=1.55in]{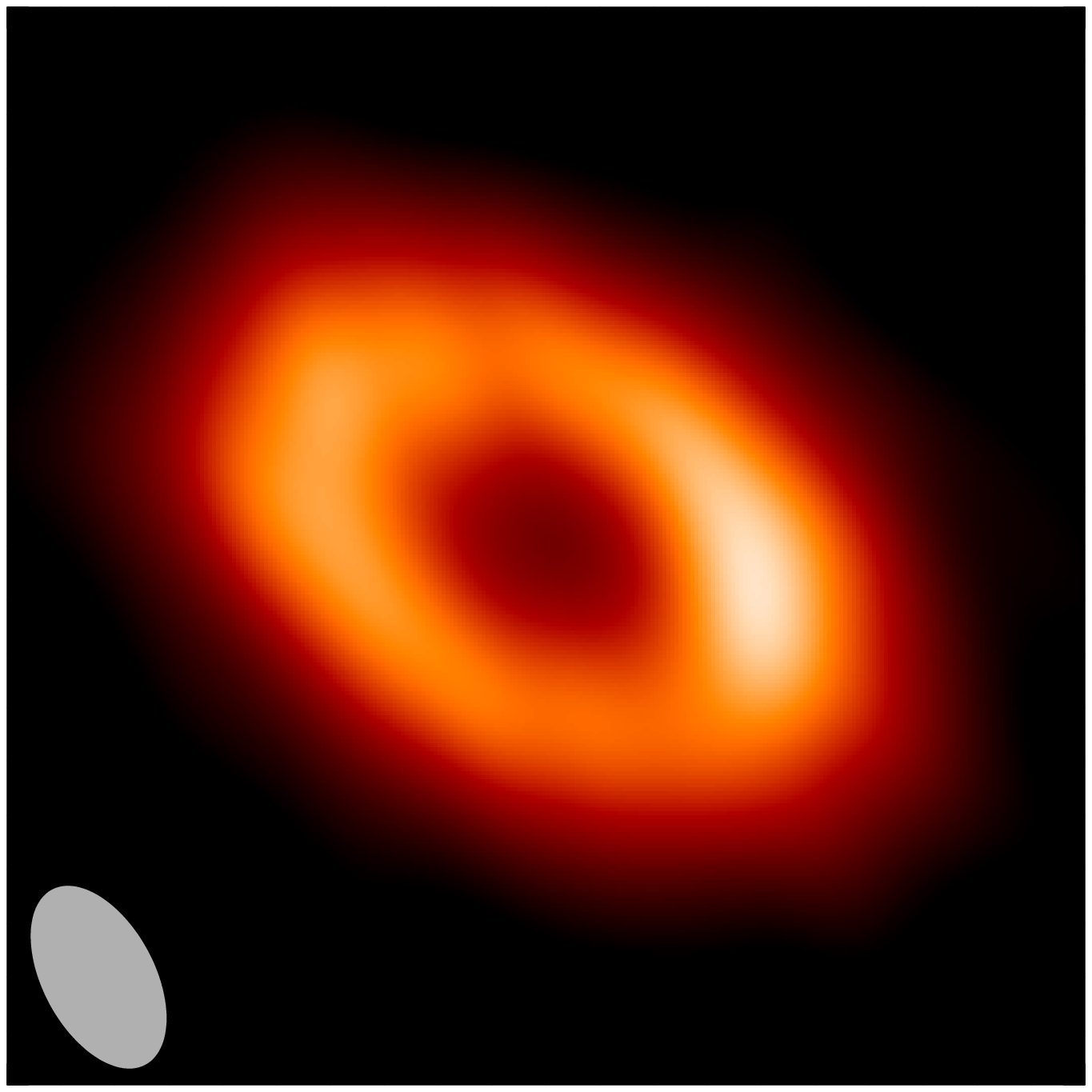} \\
\includegraphics[width=1.55in]{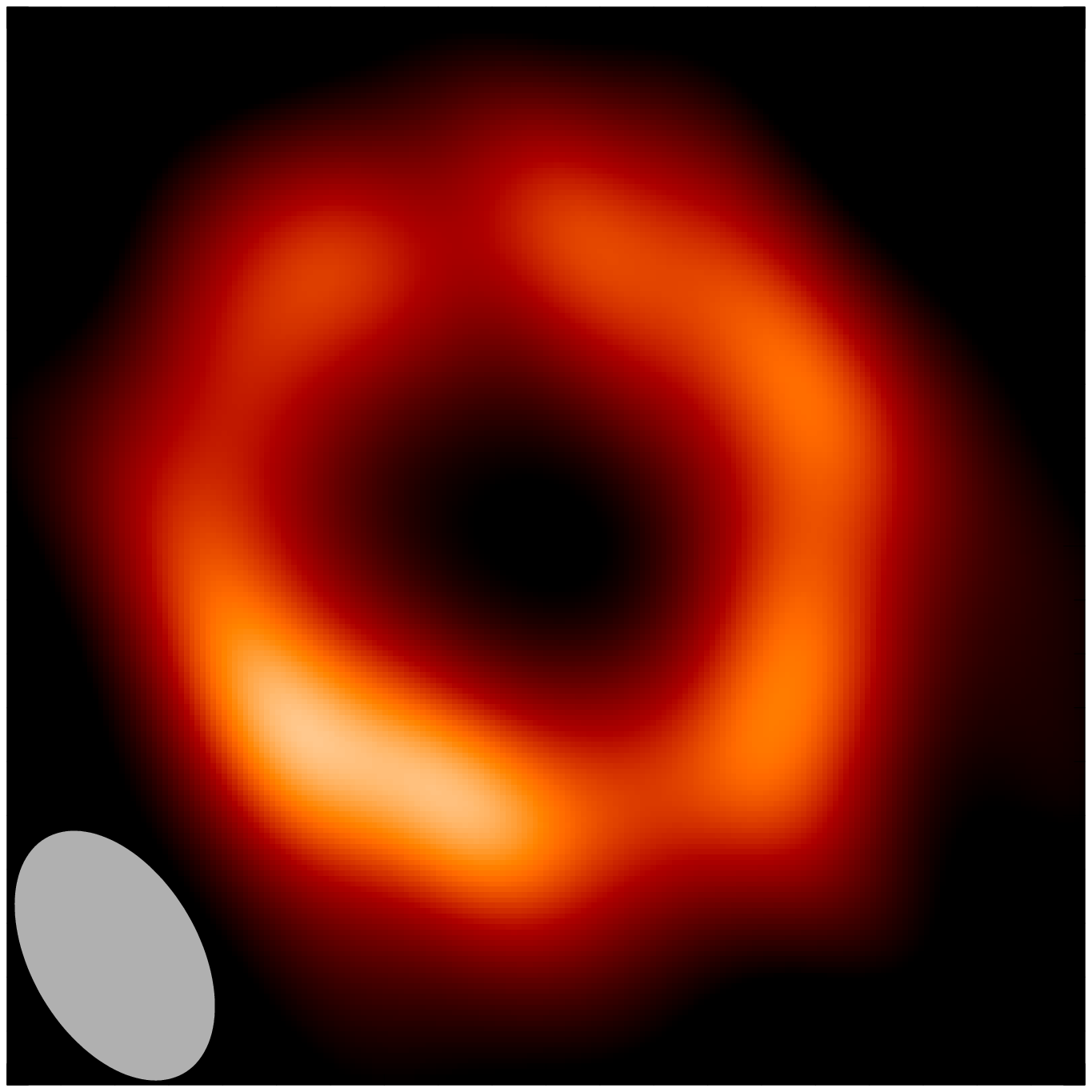} &
\includegraphics[width=1.55in]{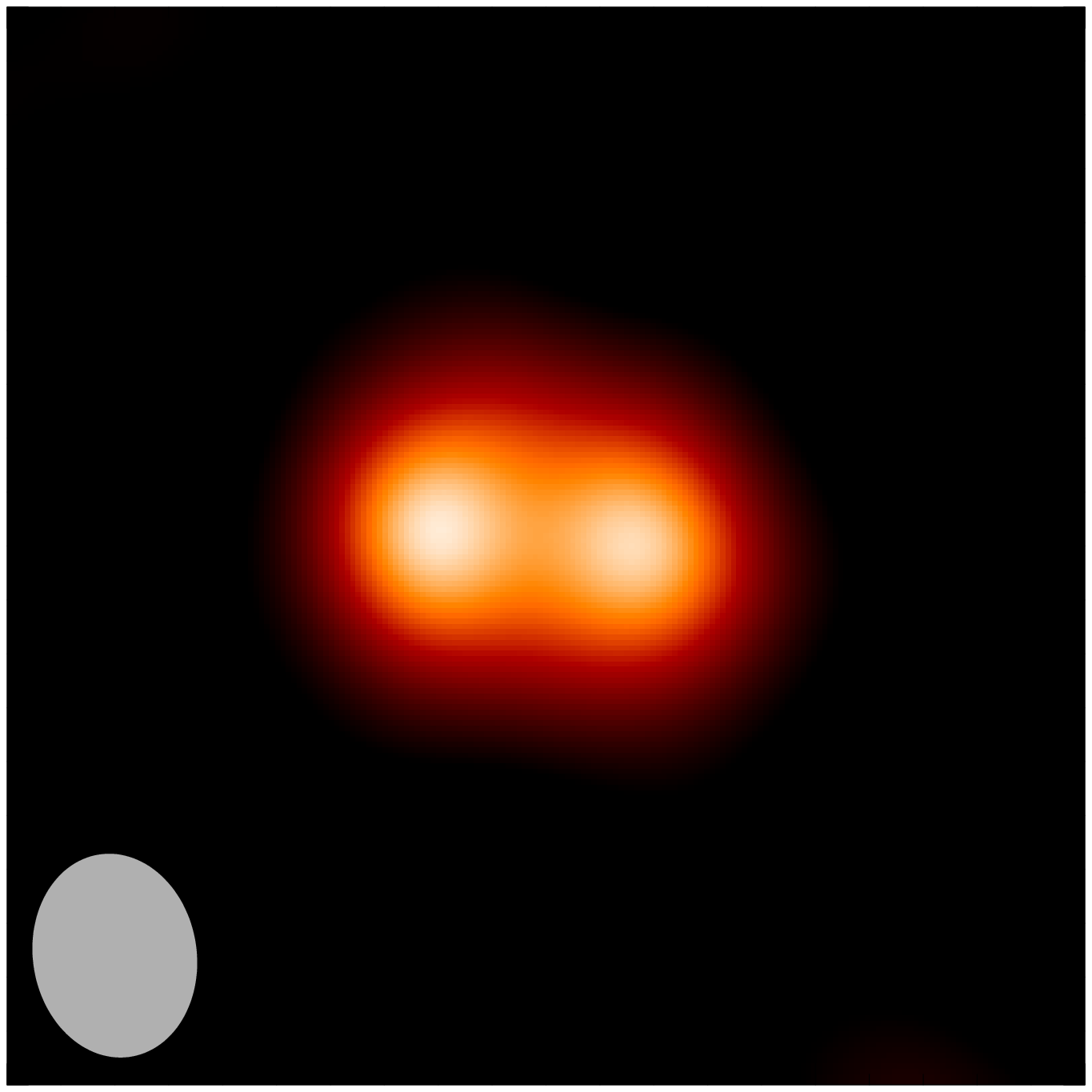} 
\end{array}$
\end{center}
\vspace{-0.1cm}
\figcaption{Examples of the 870\,$\mu$m thermal continuum ring morphologies 
observed for transition disks (at various inclinations) for transition disks: 
({\it clockwise, from top left} GM Aur \citep{hughes09}, LkCa 15 
\citep{andrews11},  RX J1633.9$-$2442 \citep{cieza12}, and RX J1604.3$-$2130 
\citep{mathews12}.  The synthesized beams are shown in the lower left of each 
panel.  
\label{fig:trans_ims}}
\end{figure}

Moreover, the ``dust ring" structures implied by these distinctive spectral 
morphologies have been directly confirmed, and characterized in detail, using 
resolved observations of the mm continuum \citep[e.g.,][]{pietu05,pietu06,
hughes07,hughes09,brown08,brown09,brown12,isella10,isella10b,andrews09,
andrews10b,andrews11,cieza12,mathews12,casassus13,fukagawa13,rosenfeld13b,
tsukagoshi14,osorio14,huelamo15,canovas15}.  Some illustrative examples are 
shown in \Cref{fig:trans_ims} (as well as the insets in 
\cref{fig:trans_rings}).  Resolved studies of transition disks have so far used
haphazard selection, and are severely limited by luminosity and resolution 
constraints.  
The current sample shows cavities with radii of $\sim$15--100\,AU and emission 
levels suppresed by at least a factor of $\sim$100, surrounded by bright, 
narrow rings (not well-resolved; potentially with high optical depths).  These 
morphological features are surprisingly more common than the infrared-based 
transition disk phenomenon in general, representing a quarter or more 
of the disks in the bright half of the mm luminosity (disk mass) function 
\citep{andrews11}.  It is not yet clear how much this apparent discrepancy is 
influenced by the strong selection biases (i.e., bright sources only) in the 
resolved mm continuum sample.

The simple ``dust ring+cavity" model for a transition disk structure becomes  
more complicated when additional tracers are considered.  The presence of a
small infrared excess (and often silicate emission features) in transition 
disk SEDs demonstrates that the cavities are not empty; rather, a ``gapped" 
structure, with an additional band of small dust grains located near the host 
star, may be a more appropriate description \citep[e.g.,][]{espaillat07,
espaillat08,espaillat10}.  Since even a low mass of small particles can emit 
strongly in the infrared, such structures can even wash out the characteristic 
SED ``dip" signature \citep[e.g.,][]{isella10,isella10b,andrews11}.  Moreover, 
spatially resolved observations of infrared scattered light (\citealt{dong12,
hashimoto12,follette13,rapson15}; see \cref{fig:trans_rings} inset) and sub-mm 
emission \citep[e.g.,][]{pinilla15} find smaller particles interior to the mm 
continuum rings, in some cases nearly filling the cavities. 

These trace populations of small particles are clearly accompanied by gas.  The 
accretion rates onto the transition disk hosts tend to be comparable to or 
slightly below those for more typical disks \citep{najita07,najita15,fang09,
espaillat12,ingleby14,manara14}, suggesting that the mass flow is relatively 
uninterrupted by the gap(s) or cavity.  Supplementary signatures of this gas 
reservoir are found in infrared rovibrational lines of common molecules located 
at few AU-scales \citep[e.g.,][]{salyk09}, unresolved emission in the wings of 
millimeter rotational lines \citep{rosenfeld12}, as well as direct spectral 
imaging of those same lines \citep{casassus13,fukagawa13,bruderer14,zhang14,
perez15,canovas15,vandermarel15}.  As with the small grains, constraints on the 
gas densities in the cavity are rudimentary.  The bulk of the gas mass remains 
confined to the extended structures that reach well-beyond the mm continuum 
rings \citep{hughes09,rosenfeld13b,walsh14}.  

Considering all these observations together, there is strong empirical evidence 
supporting the theoretical model that transition disk structures are largely 
controlled by a narrow, radial (ring-like) gas pressure maximum located outside 
a zone of substantially depleted densities.  For the specific examples that 
have been studied in detail so far, the measured accretion rates and cavity 
contents indicate that the pressure gradient outside the cavity is not too 
steep.  Such properties are generally inconsistent with the predictions of 
photoevaporation models \citep{alexander09,owen12b}, and so the operating 
hypothesis is that the observed structures are produced by dynamical 
interactions with very faint \citep[e.g.,][]{pott10,kraus11}, presumably 
planetary-mass, companions (although more massive stellar companions must be 
ruled out; e.g., \citealt{ireland08}).  That is certainly an exciting prospect, 
and is naturally driving rapid developments in the pursuit for details on these 
specific disks.  A few putative planetary-mass candidates have even been 
identified inside transition disk cavities \citep[e.g.,][]{huelamo11,kraus12b,
reggiani14,biller14,quanz14}.  That said, it is important to recognize that the 
sub-population of disks receiving most of the attention is still biased (in 
terms of mm luminosity, cavity size, etc.): a larger swath of the disk 
population may have its properties shaped more by other evolutionary factors 
(e.g., photoevaporation).

\begin{figure}[t!]
\epsscale{1.15}
\plotone{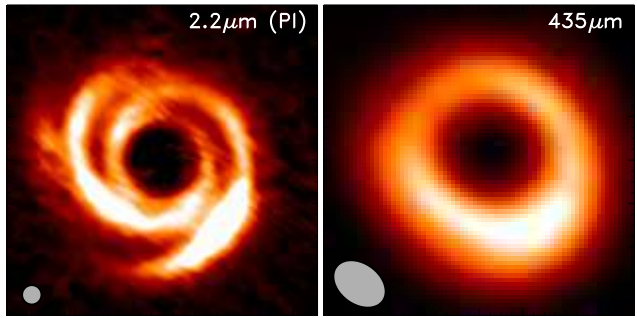}
\figcaption{An especially dramatic example of substructure in the SAO 206462
transition disk.  ({\it left}) The 2.2\,$\mu$m (polarized) scattered light 
shows prominent spiral structure \citep{garufi13}.  ({\it right}) The thermal
continuum emission at 435\,$\mu$m reveals a strong azimuthal asymmetry in the 
ring, which may be related to an unresolved spiral pattern \citep{perez14}.  
Both images are on the same scale (200\,AU on a side), with resolutions on the 
lower left.
\label{fig:asym}}
\end{figure}

\subsubsection{Substructure}

A particularly active topic of research on these disks is the characterization 
of their {\it substructure}, deviations from the simple one-dimensional 
(radial) description outlined above.  From a phenomenological perspective, 
three basic types of substructure are observed in transition disks: azimuthal 
asymmetries, spirals, and warps.  A prominent example (for two of these) is 
shown in \Cref{fig:asym}.  The former type is common in the limited samples 
available, with the millimeter emission rings in some cases appearing lopsided: 
azimuthal contrast levels range from modest (a factor of $\lesssim$2; 
\citealt{brown09,isella13,rosenfeld13b,perez14}) to extreme (a factor $>$30; 
\citealt{casassus13,vandermarel13,fukagawa13}), and the asymmetries themselves 
can account for $\sim$30--100\%\ of the integrated flux.  These asymmetric 
features are 
azimuthally resolved (spanning tens of degrees), but radially narrow 
($\lesssim$10-20\,AU).  The currently favored explanation for such substructure 
is the concentration of solids in an azimuthal gas pressure enhancement 
\citep{birnstiel13}, presumably driven by vortex instabilities produced 
exterior to the orbit of a perturbing companion \citep[e.g.,][]{barge95,
klahr97,wolf02,regaly12,lyra13,zhu14,zhu14b}.  

However, vortices are not the only viable explanations for the observed 
asymmetries; in some cases, they may alternatively be generated by global 
gravitational modes \citep{mittal15}, eccentricity \citep{kley06}, or 
unresolved spiral structures \citep[e.g.,][]{perez14}.  The latter is of 
special interest, since remarkable spiral patterns have been observed in 
(usually polarized) scattered light from many of the transition disks with 
early-type hosts (a sensitivity-related selection bias).  Both well-ordered, 
open spirals \citep[e.g.,][]{muto12,grady13,garufi13} and more complex, 
tightly-wound patterns \citep[e.g.,][]{clampin03,fukagawa04,fukagawa06,
canovas13,avenhaus14} have been identified, typically extending from the inner 
working angles ($\sim$10\,AU) out to $\gtrsim$100\,AU scales.  Some debate 
remains about the origins of these patterns, as true density waves 
\citep{dipierro14,dipierro15} or manifestations of vertical substructure 
\citep{juhasz15}, but the forthcoming resolution improvements at millimeter 
wavelengths should settle the issue.  Some care has to be taken in interpreting 
scattered light images, since the illumination pattern of the outer disk 
surface can also be affected by warped geometries \citep[e.g.,][]{roberge05,
quillen06,hashimoto11,marino15}.  Resolved spectral line data bolsters the 
evidence that such vertical substructure may be a common feature in transition 
disks \citep[e.g.,][]{rosenfeld12,rosenfeld14,perez15}.  

With new access to ALMA and high quality (polarized) scattered light data, the 
past few years have seen rapid development in resolved studies of transition 
disks.  The new signatures of dynamical substructure in these disks are 
marshaling a major investment in both observations and theoretical work.  
Ultimately, the goal is to provide firm, predictive links between the observed 
disk structures and the properties of the planetary perturbers that shape them, 
with results that have fundamental impacts on models of early planetary system 
architectures, migration, and the planetary accretion process.  At the same 
time, it is beneficial to understand that the ``sample" being studied so 
extensively is certainly biased: after all, these accreting, mm-bright 
disks with large and partially-filled cavities are but a small sub-group of the 
general transition disk population \citep[e.g.,][]{cieza08}.  It will be 
interesting to see how the description of transition disk structures changes 
when a more diverse sample is analyzed using data of similar quality.

\subsection{Synopsis}

\begin{itemize}

\item Disk solids grow and migrate toward the local maximum in the gas pressure 
distribution.  The result is a vertically and radially size-sorted density 
structure, with higher concentrations of larger particles at the midplane and 
in the inner disk.  The vertical stratification can be observed in the infrared 
SED, or through multiwavelength resolved scattered light images.  The radial 
segregation is noted in the spatially resolved mm/radio continuum ``colors".

\item Additional evidence for the inward migration of particles due to radial 
drift is available in the form of a steep drop in the dust-to-gas ratio, 
observationally manifested as a sharp-edged continuum profile and a substantial 
size discrepancy between resolved tracers of gas and mm/cm particles.  

\item The population of ``transition disks" is characterized by a ring-like 
dust geometry, with larger particles (observed in the mm/cm continuum) 
concentrated in a narrow annulus outside a central, depleted cavity.  In the 
specific examples studied in detail so far, gas and smaller particles extend 
into the cavity and out to larger radii.  The physical interpretation of these 
structures is related to the presence of a radial maximum in the gas pressure 
profile, which effectively traps large particles.  

\item A subset of the transition disk population also exhibits a variety of 
substructures, in the forms of azimuthal asymmetries, spiral patterns, and 
vertical warps.  The current thinking is that these massive, still-accreting 
systems represent young (giant) planetary systems during their formation epoch.

\end{itemize}

\noindent {\it Additional Reading}: the sequence of reviews on the evolution of 
disk solids by \citet{beckwith00}, \citet{natta07}, and \citet{testi14}; the 
more general review on disk evolution by \citet{alexander08}; and the recent 
overviews of disk dispersal and transition disks by \citet{alexander14} and 
\citet{espaillat14}.

\section{Future Directions} \label{sec:future}

The intentions of this review were to highlight some of the fundamental aspects 
of this field and their key applications, and to hopefully serve as a 
pedagogical resource (or starting point) for those interested in joining the 
related research efforts.  The past few years have seen dramatic advances in 
the observational study of circumstellar disks, and particularly their 
connections to the formation epoch for planetary systems.  It is not a stretch 
to argue that at this stage the subject could be better characterized as 
``observational planet formation".  The field is poised for substantial 
continued development, owing particular to the capabilities of the 
nearly completed ALMA project.  Given that promise, there seems little doubt 
that the landscape of disk research will look significantly different in the 
next decade.  Although it is difficult to predict that future, the first steps 
along the path toward it are now relatively clear.

Prominent among these is an effort to sort out some of the basic demographics 
of disk properties, and thereby the key factors that impact planet formation 
efficiencies.  \Cref{sec:demographics} introduced the early efforts along these 
lines, pointing out the preliminary indications that disk masses (and thereby 
the likelihood for planet formation) depend on the stellar host mass 
(\hyperref[subsec:mdms]{Sect.~\ref*{subsec:mdms}}), multiplicity 
(\hyperref[subsec:multi]{Sect.~\ref*{subsec:multi}}), and environment 
(\hyperref[subsec:enviro]{Sect.~\ref*{subsec:enviro}}).  But so far the 
limitations on the sizes and properties of samples has stifled further inquiry 
along these lines.  That will change with ALMA; several survey projects are 
underway, and more will surely follow.  There are many opportunities to address 
some fundamental questions in the field with creative explorations of the data 
available from such surveys.  How does the disk mass distribution evolve?  What 
are the inter-dependences between and evolutionary behavior of the demographic 
trends that have been noted so far?  What accounts for the tremendous scatter 
around these trends?  How does the gas reservoir, as traced by key spectral 
line ratios (e.g., \citealt{williamsbest14}), behave with respect to the 
solids?  How are these simple, compound diagnostic relations tied to more 
detailed, elemental (i.e., resolved) measurements?

The last of these questions connects back to the fundamental goal of 
constraining disk density structures (\Cref{sec:structure}).  In the near term, 
progress toward that goal from continuum data alone is a major challenge.  
Intrinsic uncertainties in the opacities of disk solids 
(\hyperref[subsec:material]{Sect.~\ref*{subsec:material}}), coupled with the 
complex processes of growth and migration that modify their sizes and spatial 
distributions (\hyperref[subsec:solidevol]{Sect.~\ref*{subsec:solidevol}}), 
make a robust and quantitative characterization of the densities prohibitive 
without additional information.  One way around this impasse is to forge closer 
links to the wealth of spectral line measurements available from ALMA 
observations.  At this stage, developing a robust forward-modeling approach to 
``fit" a large and complex union of datasets with an ever-expanding (and not 
well-justified) set of (pseudo-)physical parameters is intractable.  Instead, 
emphasis should be placed on identifying and characterizing patterns among 
resolved, multi-tracer diagnostics.  For spectral line data, these might 
include the locations of condensation fronts \citep{qi11,qi13,mathews13}, 
molecular abundance patterns (e.g., \citealt{oberg10,oberg11}; see 
\citealt{dutrey14}), or constraints on the vertical temperature distribution 
\citep{dartois03,rosenfeld13a,degregorio-monsalvo13}.  How do these and other 
measurements behave as a function of basic demographic properties?  How are 
they related to the resolved morphologies of the continuum emission?

\begin{figure}[t!]
\epsscale{1.1}
\plotone{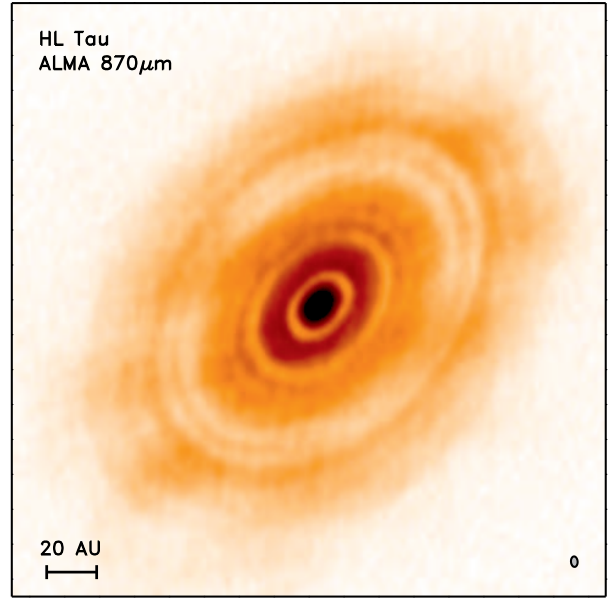}
\figcaption{The HL Tau disk observed at high angular resolution (0\farcs05, or 
$\sim$7\,AU) in the 870\,$\mu$m continuum.  The series of concentric rings are 
presumably responsible for slowing or halting radial drift, and are likely the 
active, telltale signatures of planetary system formation \citep{brogan15}.
\label{fig:hltau}}
\end{figure}

An important complementary approach involves building on the patterns already 
emerging from resolved continuum datasets, and associating these with the 
theoretical predictions from models of the evolution of disk solids 
(\hyperref[subsec:solidevol]{Sect.~\ref*{subsec:solidevol}}).  Does the 
tentative correlation between continuum luminosities and sizes 
(\hyperref[subsec:radial]{Sect.~\ref*{subsec:radial}}) hold up for larger and 
different samples, and does it depend on other demographic properties?  How is 
the radial variation in mm/radio ``color" or the location and shape of the 
outer edge of the continuum emission 
(\hyperref[subsec:solidevol]{Sect.~\ref*{subsec:solidevol}}) connected to 
factors like the evolutionary state, stellar host, or the structure of the gas 
reservoir?  And most pressing, can high angular resolution continuum 
measurements find and characterize evidence for the small-scale substructure 
presumed to be responsible for mitigating high radial drift rates, and thereby 
preserving large particles in the outer disk?  The stunning recent ALMA image 
of the canonical HL Tau disk, shown in \Cref{fig:hltau}, hints that a major 
conceptual evolution related to this last question is underway 
\citep{brogan15}.  

Substructure observed in these disks will be especially important in developing 
a comprehensive model for the formation and early evolution of planetary 
systems.  If HL Tau is any indication, observations with ALMA may find the 
general disk population riddled with young multi-planet systems carving out 
concentric, apparently resonant, gaps in their natal material.  Or perhaps the 
larger-scale asymmetries identified for many transition disks 
(\hyperref[subsec:trans]{Sect.~\ref*{subsec:trans}}) indicate that a less 
orderly phenomenology will be more common.  In any case, the underlying issue 
on this topic is finding ways to {\it quantitatively} link the small-scale 
features that are observed to the physical properties of associated planets and 
the mechanics of the disk--planet interactions.  How can multi-tracer 
measurements of (gas and dust) substructure be optimally combined to robustly 
estimate a perturber mass and differentiate it from the effects of viscous 
turbulence?  What level of diversity in scales and amplitudes are present in 
this substructure, and how does it vary as a function of the evolutionary 
state or basic demographic properties?  How can disk observations be exploited 
to optimize searches for young planets? And, when successful, what can the 
combined measurements of planets and their birth reservoirs say about the 
processes of planetary formation, migration, and accretion?

In the larger context of piecing together our cosmic origins, these are deep, 
pressing questions.  It is remarkable to think that the key technological 
resources needed to address these questions are now available, and that answers 
(and new questions) are sure to follow.  

\acknowledgments I am very grateful to David Wilner and Til Birnstiel for many 
helpful discussions and assistance related to the organization and content of 
this review.  The presentation also greatly benefitted from the insightful 
comments and suggestions kindly offered by Megan Ansdell, Xuening Bai, John 
Carpenter, Antonella Natta, Karin {\"O}berg, Chunhua Qi, Anjali Tripathi, 
Jonathan Williams, and an anonymous reviewer.  I would especially like to thank 
Til Birnstiel, Kate 
Follette, Antonio Garufi, Rita Mann, Laura P{\'e}rez, and Karl Stapefeldt for 
their willingness to share data to help make some of the figures, as well as 
John Debes, Carol Grady, Christophe Pinte, Massimo Robberto, and Glenn 
Schneider for kindly offering permission to reproduce some of their 
published figures.

\bibliography{references}
\end{document}